\documentclass[twocolumn]{aastex631}

\usepackage{threeparttable}
\usepackage{multirow}
\usepackage{hyperref}

\shorttitle{Observations for SNR G335.2+0.1 and HESS J1626$-$490}
\shortauthors{Oka et al.}

\begin{document}

\title{
Resolving the origin of the unidentified TeV source HESS J1626$-$490 as a relic of the ancient cosmic-ray factory SNR G335.2+0.1
}

\correspondingauthor{Tomohiko~Oka}
\email{toka@fc.ritsumei.ac.jp}

\author[0000-0002-9924-9978]{Tomohiko~Oka}
\affiliation{Department of Physical Sciences, Ritsumeikan University, Kusatsu, Shiga 525-8577, Japan}
\affiliation{Research Organization of Science and Technology, Ritsumeikan University, Kusatsu, Shiga 525-8577, Japan}

\author[0000-0002-7005-7139]{Wataru~Ishizaki}
\affiliation{Center for Gravitational Physics, Yukawa Institute for Theoretical Physics, Kyoto University, Kyoto, Kyoto 606-8502, Japan}
\affiliation{Astronomical Institute, Graduate School of Science, Tohoku University, Sendai 980-8578, Japan}

\author[0000-0003-2921-1592]{Masaki~Mori}
\affiliation{Department of Physical Sciences, Ritsumeikan University, Kusatsu, Shiga 525-8577, Japan}

\author[0000-0003-2062-5692]{Hidetoshi~Sano}
\affiliation{Faculty of Engineering, Gifu University, 1-1 Yanagido, Gifu 501-1193, Japan}

\author[0000-0002-8152-6172]{Hiromasa~Suzuki}
\affiliation{Faculty of Engineering, University of Miyazaki, Miyazaki 889-2192, Japan}

\author[0000-0002-4383-0368]{Takaaki~Tanaka}
\affiliation{Department of Physics, Konan University, 8-9-1 Okamoto, Higashinada, Kobe, Hyogo 658–8501, Japan}

\begin{abstract}

While decades of observations in the TeV gamma-ray band have revealed more than 200~sources with radio or X-ray counterparts, there remain dozens of unidentified TeV sources, which may provide crucial information of cosmic ray (CR) accelerators.
HESS J1626$-$490 is an unidentified TeV gamma-ray source but is expected to originate from CRs that escaped from the nearby supernova remnant (SNR) G335.2+0.1 and are interacting with dense interstellar clouds.
To test this scenario, we scrutinize the properties of the SNR and search for non-thermal counterparts by analyzing observational data in the radio, X-ray, and GeV gamma-ray bands.
From analysis of the H\,{\sc i} and $^{12}$CO ($J{=}1{-}0$) line data, we identify the cloud associated with the SNR and compare the morphologies of the cloud and the
gamma-ray emission.
The distance and age of the SNR are estimated to be $3.3 \pm 0.6$~kpc and ${\sim}5$~kyr, respectively.
From X-ray and GeV gamma-ray data analyses, we find an extended GeV gamma-ray emission overlapping with the SNR and H.E.S.S. source regions but no X-ray emission.
The location of the peak of the extended GeV emission changes from near the SNR at $\lesssim 1$~GeV to the H.E.S.S. source at $>10$~GeV.
We find a spectral hardening at ${\sim}50$~GeV, which is consistent with the existence of two components in the gamma-ray emission.
We find that a combination of emission from the escaped CRs and the SNR itself can reproduce the observed broadband spectrum, on the assumption that the SNR has accelerated protons to ${\gtrsim}100$~TeV in the past.

\end{abstract}


\section{Introduction} \label{section:Introduction}

The origin of cosmic rays (CRs) remains a mystery for more than a century after their discovery \citep{Hess1912}, while supernova remnants (SNRs) in our Galaxy are widely believed to accelerate the bulk of them \citep{Baade1934PNAS}, especially up to ${\sim}$PeV energies.
To elucidate this, many gamma-ray observations have been conducted for decades, as gamma rays are reliable probes of high-energy CRs \citep[e.g.,][for a review]{Cristofari2021}.
Recent gamma-ray observations of the Galactic diffuse emission \citep{TibetASgamma2021} and a single SNR \citep{Fang2022PhRvL, MAGIC2023A&A} have provided strong evidence for proton acceleration up to PeV in our Galaxy.
However, many unidentified gamma-ray sources have been found \citep[e.g.,][]{HESS2018, FermiLAT2019, Cao2024ApJS}, representing a puzzle to get a complete picture of CR acceleration and propagation.

\begin{figure}
    \centering
    \includegraphics[width=\hsize]{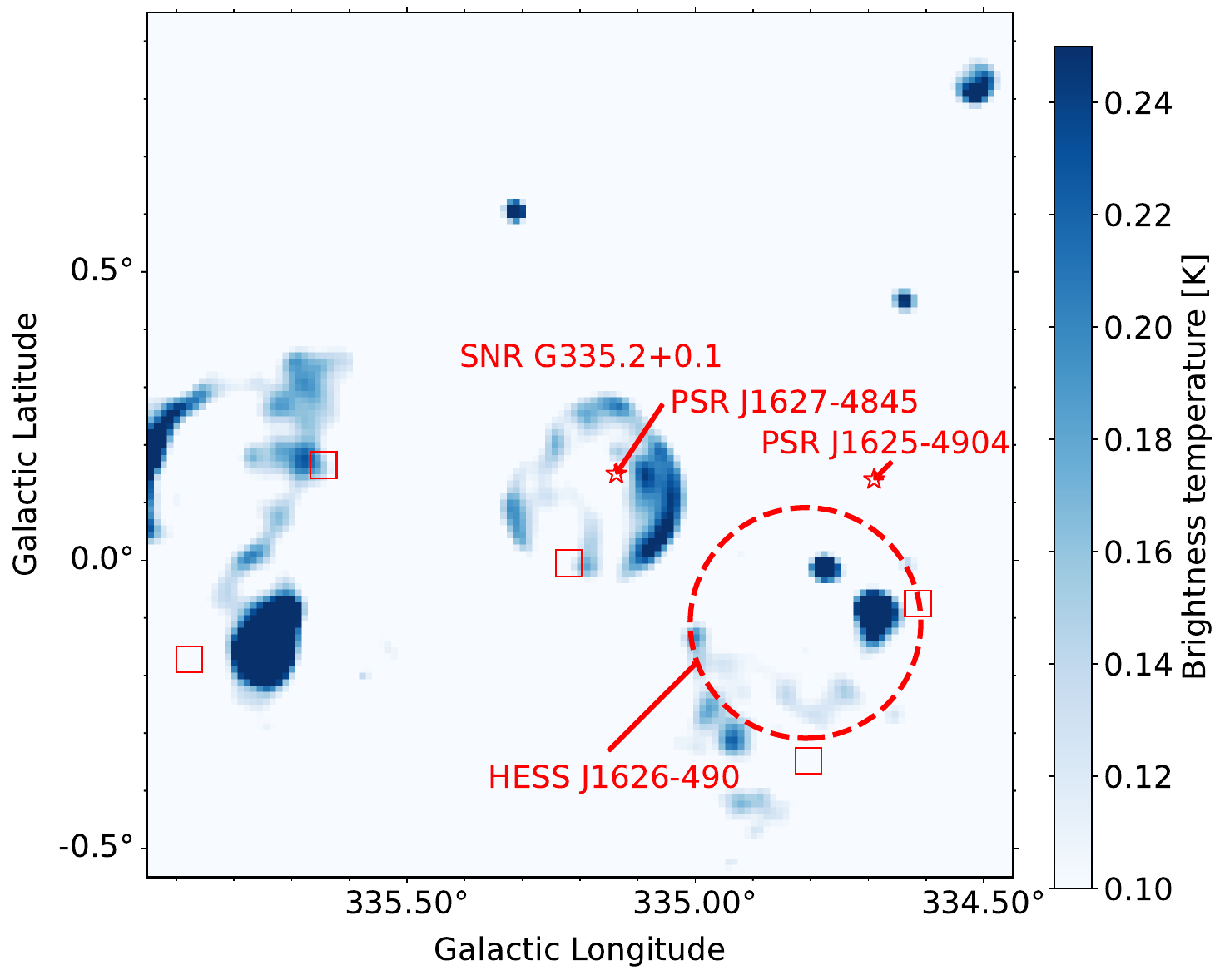}
    \caption{Radio continuum map at 1420~MHz in the vicinity of SNR G335.2$+$0.1, obtained from the SGPS data \citep{McClure-Griffiths2005}.
    The dashed circle represents the extended TeV gamma-ray emission of HESS J1626$-$490.
    Note that the two bright emissions in the circle are H\,{\sc ii} regions \citep{Russeil2003A&A, Eger2011}, which are not associated with the TeV source.
    The square markers indicate the positions of GeV gamma-ray sources listed in the 4FGL-DR3 catalog \citep{Abdollahi2022ApJS}.
    The star markers show the positions of pulsars, PSR J1627$-$4845 \citep{Kaspi1996} and PSR J1625$-$4904 \citep{Manchester2001MNRAS}.
    }
    \label{fig:radio_continuum}
\end{figure}

HESS J1626$-$490 is one of the unidentified TeV gamma-ray sources \citep{HESS2007}, which is located close to SNR G335.2$+$0.1, as shown in Fig.~\ref{fig:radio_continuum}.
By fitting with a symmetric Gaussian function, the center position of the H.E.S.S. source is estimated to be (RA, DEC) = (246.75 $\pm$ 0.07$^{\circ}$ , $-$49.17 $\pm$ 0.07$^{\circ}$) (J2000), which is offset from the location of SNR G335.2+0.1 by 0.4$^{\circ}$.
The extent of the emission is 0.200 $\pm$ 0.035$^{\circ}$.
\cite{Eger2011} analyzed radio emission line and X-ray data and then found a spatial coincidence between an H\,{\sc i} cloud and the TeV emission, while no X-ray counterpart was found.
The authors therefore attempted to explain the TeV emission by assuming
a CR-escape scenario \citep[e.g.,][]{Aharonian1996A&A, Gabici2007ApJL}, where protons accelerated at SNR G335.2$+$0.1 escape from an acceleration site and illuminate nearby clouds, which produce gamma-ray emission via the $\pi^{0}$-decay process.
Gamma-ray energy spectra at the clouds mainly depend on the diffusion coefficient, the time evolution of maximum CR energy at an SNR, target proton densities, and the distance between an SNR and cloud regions \citep{Gabici2007ApJL, Ohira2010_MNRAS, Oka2022}.
Hence, to test for the above scenario requires information on the broadband non-thermal spectrum and the environment (e.g., distance from Earth, age of the SNR, and gas density) of the SNR itself and clouds.

The SNR G335.2+0.1 has a shell-type radio morphology with an angular diameter of $0.33^{\circ}$ \citep{Whiteoak1996} and a flat spectrum with $\alpha = 0.46$ \citep{Clark1975} or $0.5$ \citep{Green2019JApA} being the index of flux density $S_{\nu} \propto \nu^{-\alpha}$.
The association of H\,{\sc i} clouds with SNR G335.2$+$0.1 suggests that the distance is 1.8~kpc \citep{Eger2011}, while the estimates from the $\Sigma$-D relation and an absorption study using red clump stars are 4.2~kpc \citep{Pavlovic2013} and $3.91 \pm 0.49$~kpc \citep{Wang2020}, respectively.
Although no counterpart is found in the X-ray band \citep{Eger2011}, the {\it Fermi}-LAT observations have revealed GeV gamma-ray emission spatially coinciding with SNR G335.2$+$0.1 and four other unidentified point sources within $1^{\circ}$ of the SNR \citep{FermiLAT2019}.
In detail, 4FGL J1626.0$-$4917c is the closest object to the H.E.S.S. source, but \cite{Hui2020MNRAS} lists it as a pulsar candidate.
\cite{Bhat2022A&A} reported that there is no clear association between these unidentified objects and active galactic nuclei.
In addition, they also note the spatial correspondence between 4FGL J1631.7$-$4826c and a dark cloud.
We also remind that 4FGL J1626.0$-$4917c, J1629.3$-$4822c, and J1631.7$-$4826c out of five sources have ’c’ in their name, indicating that they are considered to be potentially confused with galactic diffuse emission rather than point sources unrelated to the SNR or the H.E.S.S. source.
Thus, although the origin of each emission component has been discussed, there is no conclusive evidence, and the relation with the H.E.S.S. source and/or the SNR is not well understood.

In this work, we explore the association between the unidentified TeV source HESS J1626$-$490 and SNR G335.2$+$0.1 using the latest multi-wavelength observational data.
We first search for non-thermal emissions from both the H.E.S.S. and SNR regions in the GeV gamma-ray and X-ray bands.
We also analyze H\,{\sc i} and $^{12}$CO ($J=1-0$) line data to examine discrepancies between estimates from the previous H\,{\sc i} study and the other methods in the distance to the SNR.
We then test the CR-escape scenario for the H.E.S.S. source with spectral modeling.
Sect.~\ref{section:Observation} describes the details of observational data used in this study, and Sect.~\ref{section:AnalysisResults} shows our analysis results.
In Sect.~\ref{section:modeling}, we show modeling results with the broad-band spectrum.
The discussion and summary are presented in Sect.~\ref{section:Discussion} and \ref{section:Summary}, respectively.

\section{Observations} \label{section:Observation}

\subsection{GeV gamma-ray data} \label{sec:obs:Fermi}

The Large Area Telescope (LAT) onboard the {\it Fermi} mission, launched in 2008, is capable of detecting gamma rays in the energy range from ${\sim}100$~MeV to ${\gtrsim}1$~TeV \citep{FermiLAT2009}. 
We analyze its 13.5-year data from 2008 August to 2022 April in the vicinity of HESS J1626$-$490, using the standard analysis software, the latest Fermitools package v2.2.11 \citep{Fermitools2019}.
We use the ``Source'' selection criteria and instrument responses (\verb|P8R3_SOURCE_V3|\footnote{\url{https://fermi.gsfc.nasa.gov/ssc/data/analysis/documentation/Cicerone/}}), considering a balance between precision and photon statistics.
We select gamma rays with zenith angles greater than 90$^{\circ}$ to suppress the contamination of the background from the Earth limb.
We use a standard maximum likelihood method \citep{Mattox1996} for spatial and spectral analyses in the binned mode.
We choose a square region of 14 $\times$ 14$^{\circ}$ with the center coinciding with that of SNR G335.2$+$0.1, (RA, DEC) = ($246.938^{\circ}$, $-48.783^{\circ}$), as the region of interest (ROI) for the (binned) maximum likelihood analysis based on Poisson statistics.
The source spatial-distribution model includes all the sources in the fourth \textit{Fermi} catalog~\citep[4FGL-DR3;][]{FermiLAT2019, Abdollahi2022ApJS} within $30^{\circ}$ from the SNR and the two diffuse backgrounds, the Galactic (\verb|gll_iem_v7.fits|) and extragalactic (\verb|iso_P8R3_SOURCE_V3_v1.txt|) diffuse emissions.
Note that the significance of a source is represented in this analysis by the Test Statistic (TS) defined as $-2 \mathrm{log}(L_{0}/L)$, where $L_{0}$ and $L$ are the maximum likelihood values for the null hypothesis and a model including additional sources, respectively~\citep{Mattox1996}.

\subsection{X-ray data} \label{sec:obs:XMM}

An X-ray observation of the G335.2+0.1 region has been performed with XMM-Newton \citep{jansen01} EPIC MOS \citep{turner01} and pn \citep{struder01} CCD cameras (ObsID: 0724560101).
For data processing and analysis, we use the XMM-Newton Science Analysis System \citep[SAS;][]
{Gabriel2004ASPC} version 19.1.0 with Extended Source Analysis Software \citep[ESAS;][]{snowden04}.
In our spectral analysis, we use HEASoft \citep{heasarc14} version 6.20, XSPEC \citep{arnaud96} version 12.9.1, and AtomDB \citep{smith01, foster12} version 3.0.9.
After the standard data processing for extended sources with the SAS tasks {\tt pn-filter, epchain, emchain}, and {\tt mos-filter}, effective exposures are found to be 21~ks (MOS1), 22~ks (MOS2), and 19~ks (pn).

\subsection{Radio data} \label{sec:obs:Radio}

We use $^{12}$CO ($J=1-0$) line data obtained with the Mopra telescope \citep{Braiding2018PASA}, which is a 22-m single-dish radio telescope in Australia.
The angular resolution is $0\farcm6$, which is better than those of telescopes used in previous studies for this object.
The typical noise fluctuation is ${\sim}$1.2~K at a velocity resolution of 0.1~km\,s$^{-1}$.

We also use the H\,{\sc i} line and radio continuum data at 1.4~GHz, which were taken in the Southern Galactic Plane Survey~\citep[SGPS;][]{McClure-Griffiths2005} with the Australia Telescope Compact Array (ATCA) and Parkes 64~m radio telescope.
The angular resolution is $2\farcm2$, while the velocity resolution is $0.8~{\rm km\,s^{-1}}$.
The typical noise fluctuation is 1.6~K.

\section{Results} \label{section:AnalysisResults}

\subsection{{\it Fermi}-LAT data analysis} \label{sec:ana:fermi}

\begin{figure*}
    \begin{tabular}{cc}
        \begin{minipage}{0.425\hsize}
            \centering
            \includegraphics[width=\hsize]{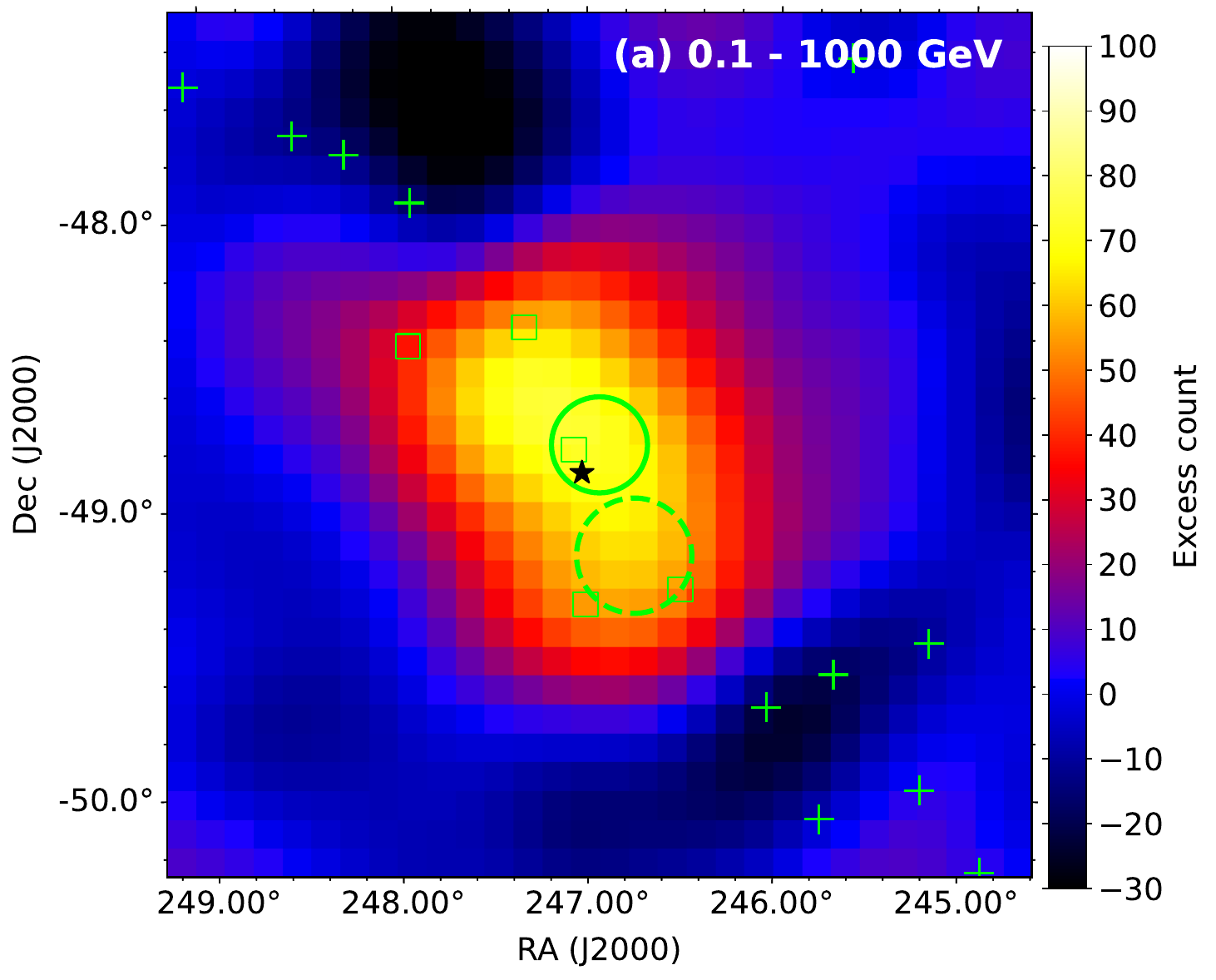}
        \end{minipage}
        &
        \begin{minipage}{0.425\hsize}
            \centering
            \includegraphics[width=\hsize]{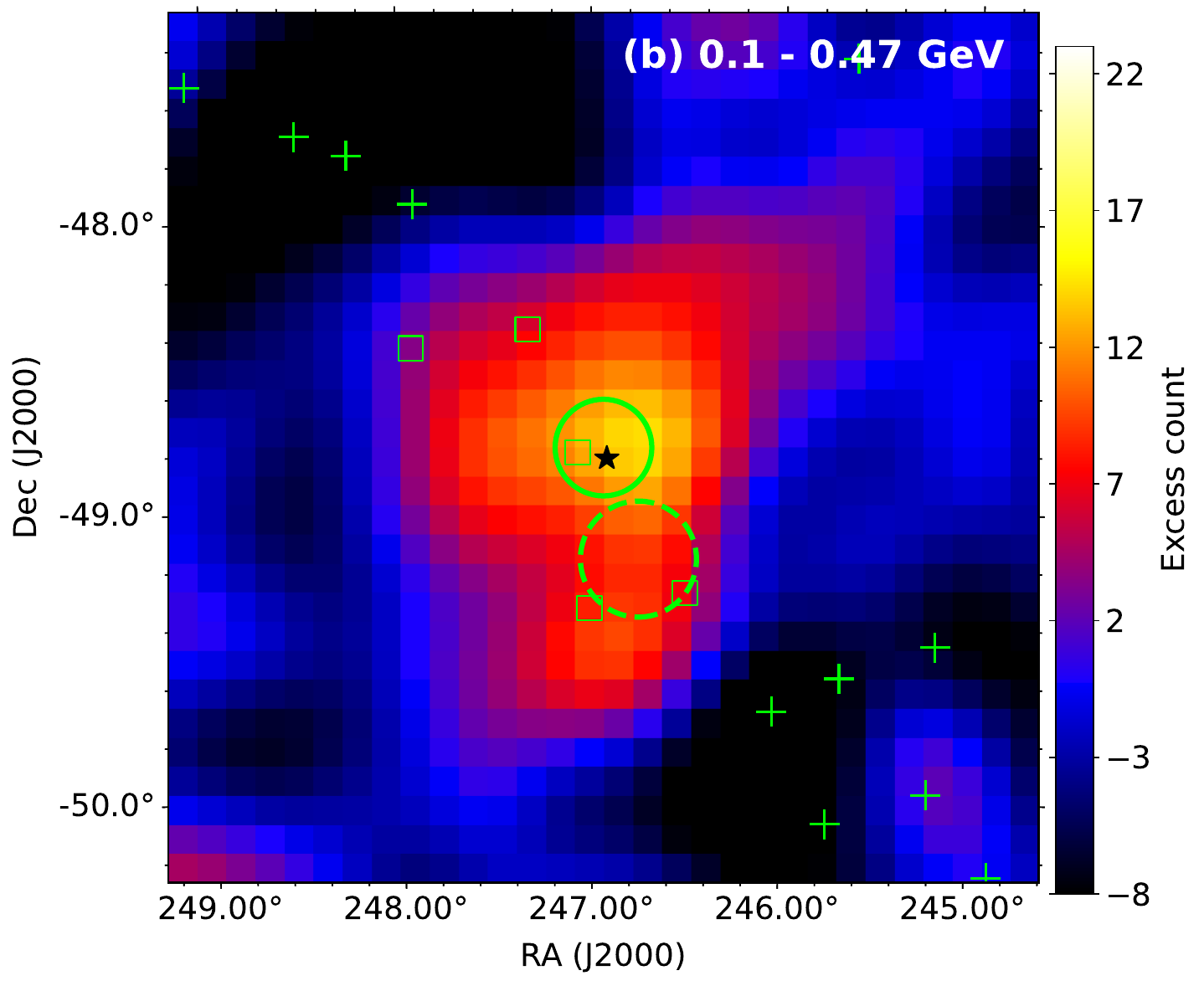}
        \end{minipage}
        \\
        \begin{minipage}{0.425\hsize}
            \centering
            \includegraphics[width=\hsize]{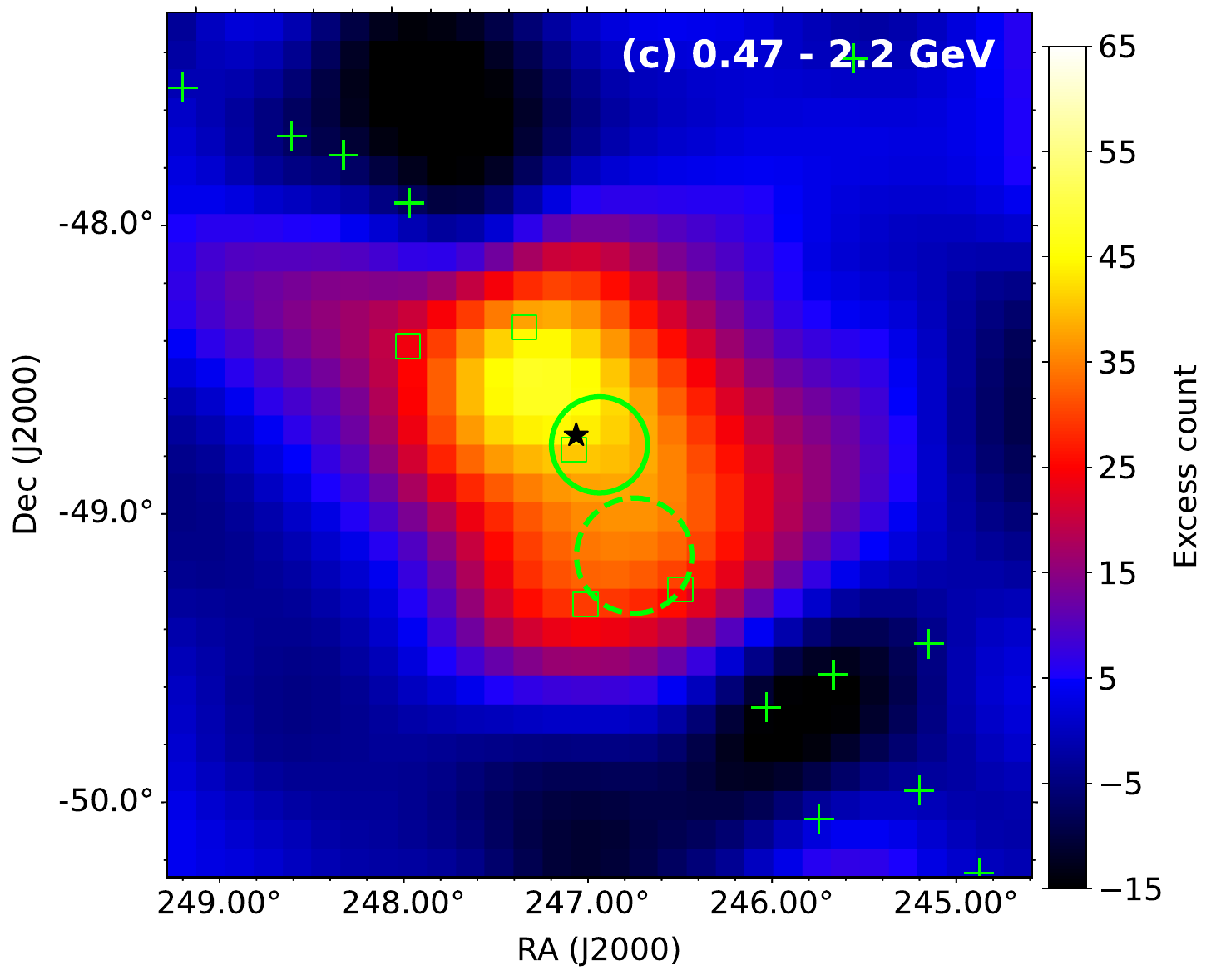}
        \end{minipage}
        &
        \begin{minipage}{0.425\hsize}
            \centering
            \includegraphics[width=\hsize]{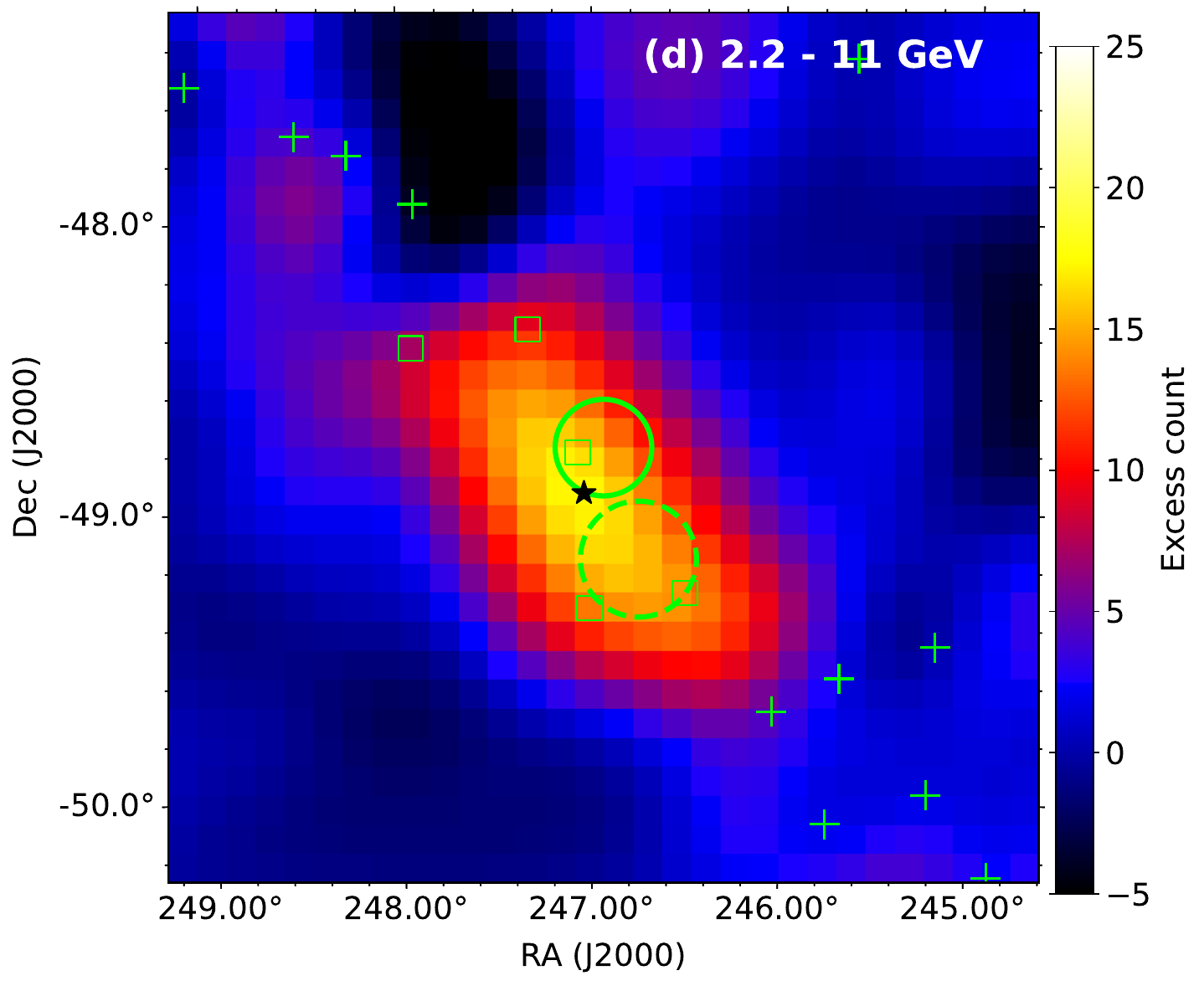}
        \end{minipage}
        \\
        \begin{minipage}{0.425\hsize}
            \centering
            \includegraphics[width=\hsize]{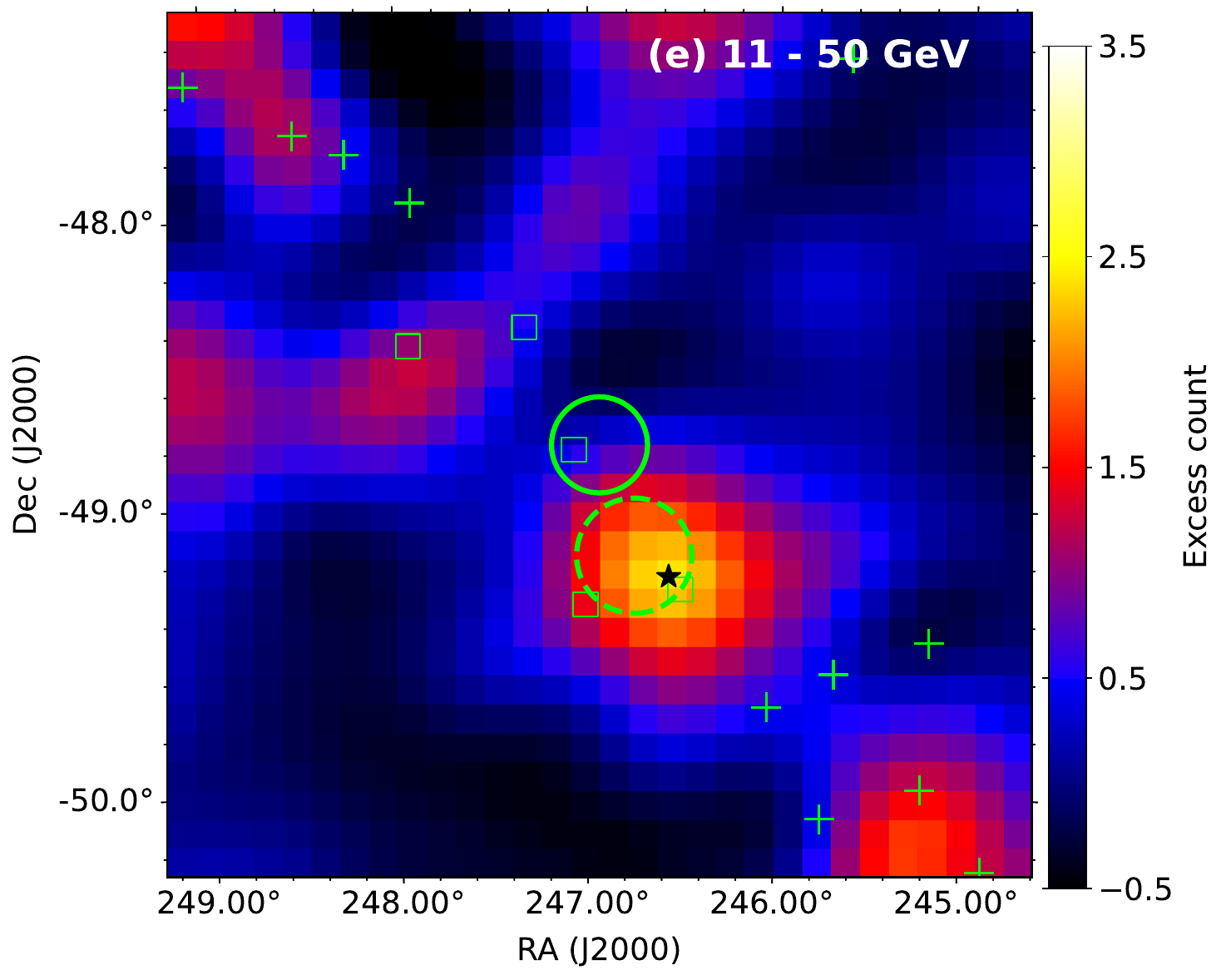}
        \end{minipage}
        &
        \begin{minipage}{0.425\hsize}
            \centering
            \includegraphics[width=\hsize]{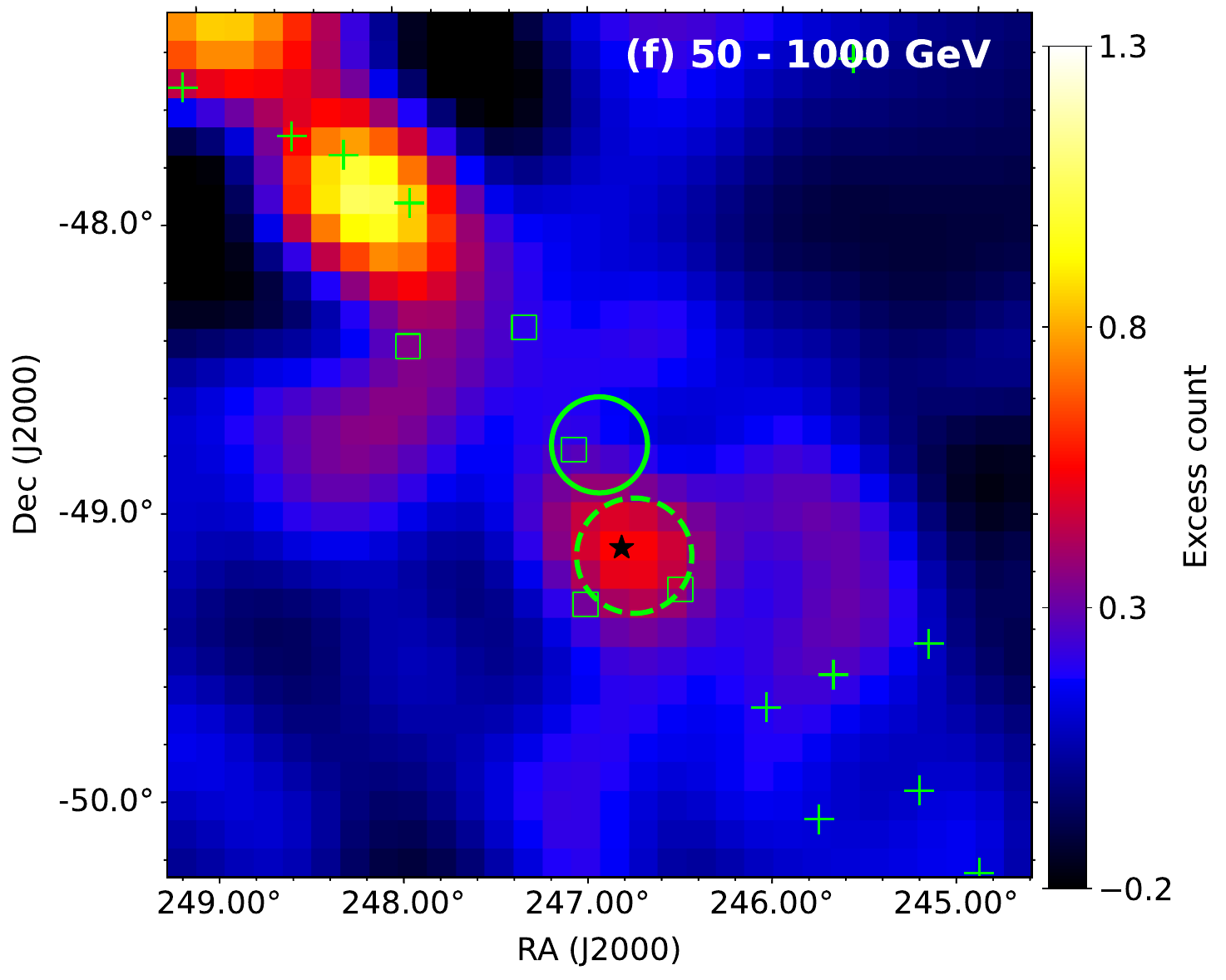}
        \end{minipage}
  \end{tabular}
\caption{Energy-dependent GeV gamma-ray skymaps in the vicinity of SNR G335.2$+$0.1, obtained from {\it Fermi}-LAT observations.
    The color scale represents the number of excess events from the background model, and all panels are smoothed by a Gaussian kernel with a radius of $0.2^{\circ}$.
    The selected energy ranges are (a) 0.1--1000~GeV, (b) 0.1--0.47~GeV, (c) 0.47--2.2~GeV, (d) 2.2--11~GeV, (e) 11--50~GeV, and (f) 50--1000~GeV, respectively.
    In each panel, the solid and dashed circles represent the location and extent of the SNR radio shell and the H.E.S.S. source, respectively.
    The green markers indicate the positions of the 4FGL-DR3 sources, of which the crosses (squares) are those included in (removed from) the background model.
    The black star shows the center position of the fitted Gaussian function in each energy band.
}
\label{fig:GeVmap}
\end{figure*}

\begin{figure}
    \centering
    \includegraphics[width=\hsize]{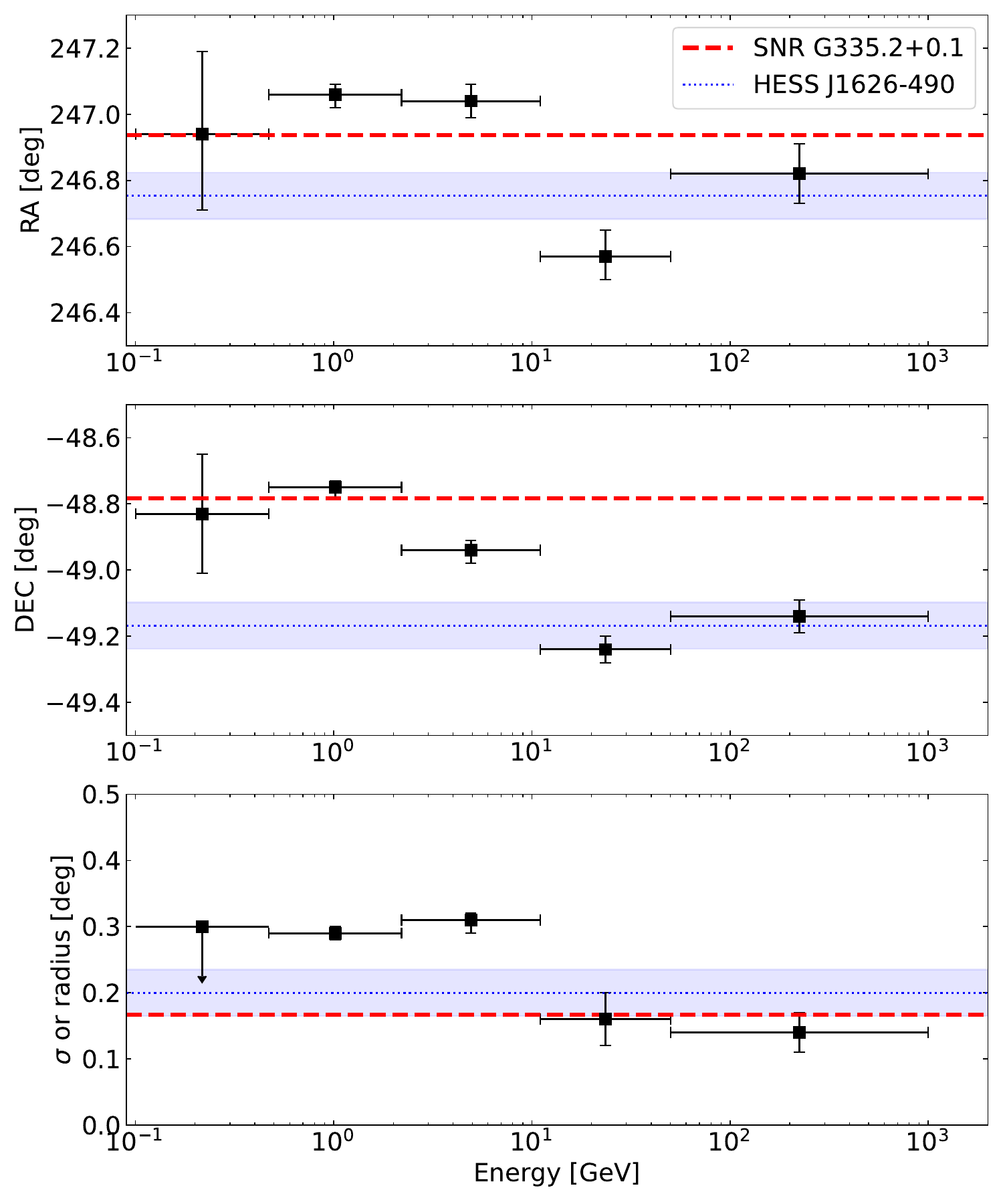}
    \caption{The fitted Gaussian parameters of the LAT $\gamma$-ray maps as a function of energy.
    The top and middle panels show the coordinates (RA and DEC) of the center position of the Gaussian, and the bottom panel shows the $1\sigma$ dispersion of the Gaussian.
    The dashed red lines represent the coordinates and radius of SNR G335.2$+$0.1, while the blue shaded areas represent those of HESS J1626$-$490 \citep{HESS2018}.    
    }
    \label{fig:Fermi_GaussFit}
\end{figure}

To identify the GeV counterpart of the H.E.S.S. source, we investigate the morphology in the GeV band using \textit{Fermi}-LAT data.
As for the background sources, we consider the Galactic diffuse emission, the isotropic background, and the 4FGL sources, as described in Sect.~\ref{section:Observation}.
However, among the 4FGL objects, five point sources within $1^{\circ}$ of the SNR are excluded from the background model because they may be related to the emission of the SNR and the H.E.S.S. source, as mentioned in Sect. \ref{section:Introduction}.
Fig.~\ref{fig:GeVmap}a shows the excess count map for 0.1--1000~GeV after the background emissions are subtracted.
An apparent excess is seen from a large area that contains both the SNR and H.E.S.S. source regions.
We evaluate the GeV morphology by fitting it with a symmetric two-dimensional (2-d) Gaussian function.
In this fitting procedure, we add a Gaussian source to the background model and search for a parameter set yielding the best likelihood value.
This likelihood fit considers the point-spread function (PSF) under the assumption that the energy spectrum of the Gaussian source follows a simple power-law function with an index of $-2$.
The fitted values are (RA, DEC) = ($247.03 \pm 0.04^{\circ}$, $-48.88 \pm 0.03^{\circ}$) (J2000) and $\sigma = 0.31\pm0.01^{\circ}$.
Panels b--f of Fig.~\ref{fig:GeVmap} show the energy-dependent morphology in the five gamma-ray bands.
We find that the location of the GeV emission gradually changes with energy.
As with the above, we fit each GeV gamma-ray map with a symmetric 2-d Gaussian function and show the results in Fig.~\ref{fig:Fermi_GaussFit}.
In the low-energy band (0.1~GeV -- 2.2~GeV), the center position is in good agreement with the SNR, while in the high-energy band ($> 11~{\rm GeV}$), the Gaussian parameters are consistent with the H.E.S.S. source\footnote{In Panel f, another excess is visible to the northeast of SNR G335.2+0.1. This area contains the two extended GeV sources FGES J1631.6$-$4756 and FGES J1633.0$-$4746, which are likely associated with the PWN HESS J1632$-$478 \citep{Fermi2017ApJ_FGES}.}.

\begin{figure}
    \centering
    \includegraphics[width=0.95\hsize]{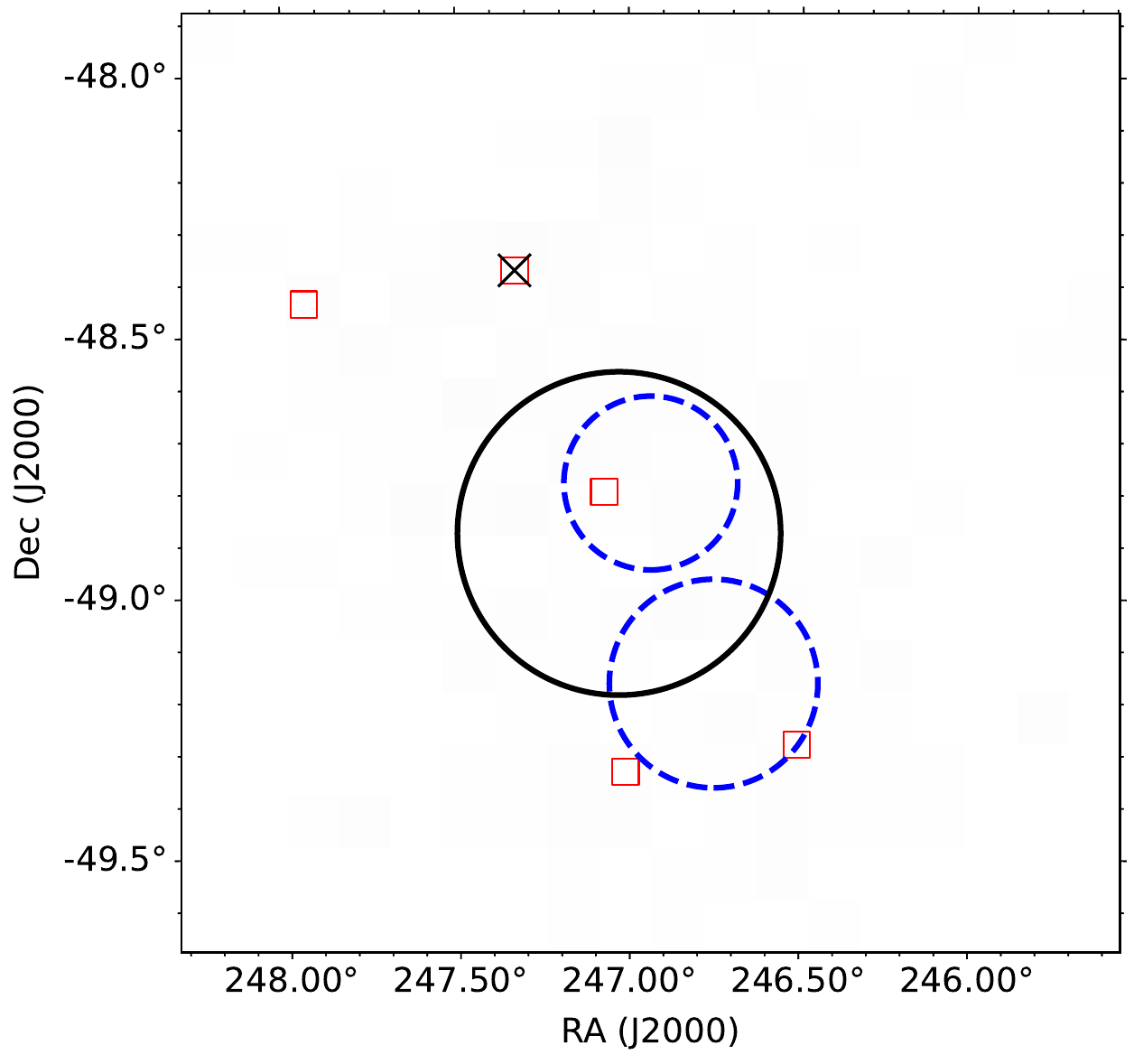}
    \caption{The schematic of the spatial models used in the likelihood test.
    The black solid and blue dashed circles indicate the locations of the Gaussian sources for the 1-Gaussian model and the 2-Gaussian model, respectively.
    The square markers show the five point sources in the 5-PS model, while the x marker shows 4FGL J1629.3$-$4822c, adopted for the 1-Gaussian+1-PS model.
    }
    \label{fig:Fermi_SpatialModel}
\end{figure}

We search for the spatial model that best describes the gamma-ray distribution by comparing likelihood-fit results in the energy range of 100~MeV--1~TeV.
In this test, the null hypothesis considers the same situation as the background model assumed above (the skymap analysis) and consists of the two diffuse backgrounds and 4FGL sources except for five point sources within 1$^{\circ}$ from the center of the SNR.
Here, we test the following three models: five point sources as suggested in the 4FGL catalog \citep{FermiLAT2019, Abdollahi2022ApJS} (``5-PS'' model), single Gaussian source (``1-Gaussian'' model), and two Gaussian sources (``2-Gaussian'' model).
As for the Gaussian parameters, we adopt the fit result (RA = $247.03^{\circ}$, DEC = $-48.88^{\circ}$, $\sigma$ = $0.31^{\circ}$) to the skymap of 0.1--1000 GeV for the 1-Gaussian model, while we employ the spatial parameters of the H.E.S.S. source (RA = $246.75^{\circ}$, DEC = $-49.17^{\circ}$, $\sigma$ = $0.20^{\circ}$; \citet{HESS2007}) and the SNR radio shell (RA = $246.94^{\circ}$, DEC = $-48.78^{\circ}$, $\sigma$ = $0.17^{\circ}$; \citet{Green2019JApA}) for the 2-Gaussian model.
For convenience, we refer to the Gaussian source in the 1-Gaussian model as Gaussian-W.
A schematic of the spatial models used in this test is given in Fig. \ref{fig:Fermi_SpatialModel}.
The likelihood-fit results with each spatial model are summarized in Table \ref{tab:likelihood}.
We find that the 5-PS model provides the best TS value among the three models, while no significant improvement is found between the 2-Gaussian model and the 1-Gaussian model.
Next, we test a model that combines Gaussian-W with a point source included in the 5-PS model (that we name ``1-Gaussian+1-PS'' model).
We then find that the model with 4FGL J1629.3$-$4822c yields the best TS value, which is better than the 5-PS model.
Considering the parameter degrees of freedom in the fit, we conclude that the combined model (1-Gaussian+1-PS model) of the one extended Gaussian source and the one point source 4FGL J1629.3$-$4822c is significantly preferred over the 5-PS model.

\begin{table}
    \centering
    \caption{Likelihood test results for spatial models using the spatial models given in Fig. \ref{fig:Fermi_SpatialModel}.
    The second and third columns indicate the difference in TS and the number of degrees of freedom from values in the null hypothesis, respectively.}
    \label{tab:likelihood}
    \begin{tabular}{lcc}
    \hline \hline
    Model & $\Delta {\rm TS}$ & $\Delta$ndf \\ \hline
    Null hypothesis & 0 & 0 \\
    5-PS & 880.1 & 13 \\
    1-Gaussian & 835.9 & 3 \\
    2-Gaussian & 831.8 & 5 \\
    1-Gaussian+1-PS & 886.9 & 6 \\
    \hline
    \end{tabular}
\end{table}

\begin{figure}
    \centering
    \includegraphics[width=0.95\hsize]{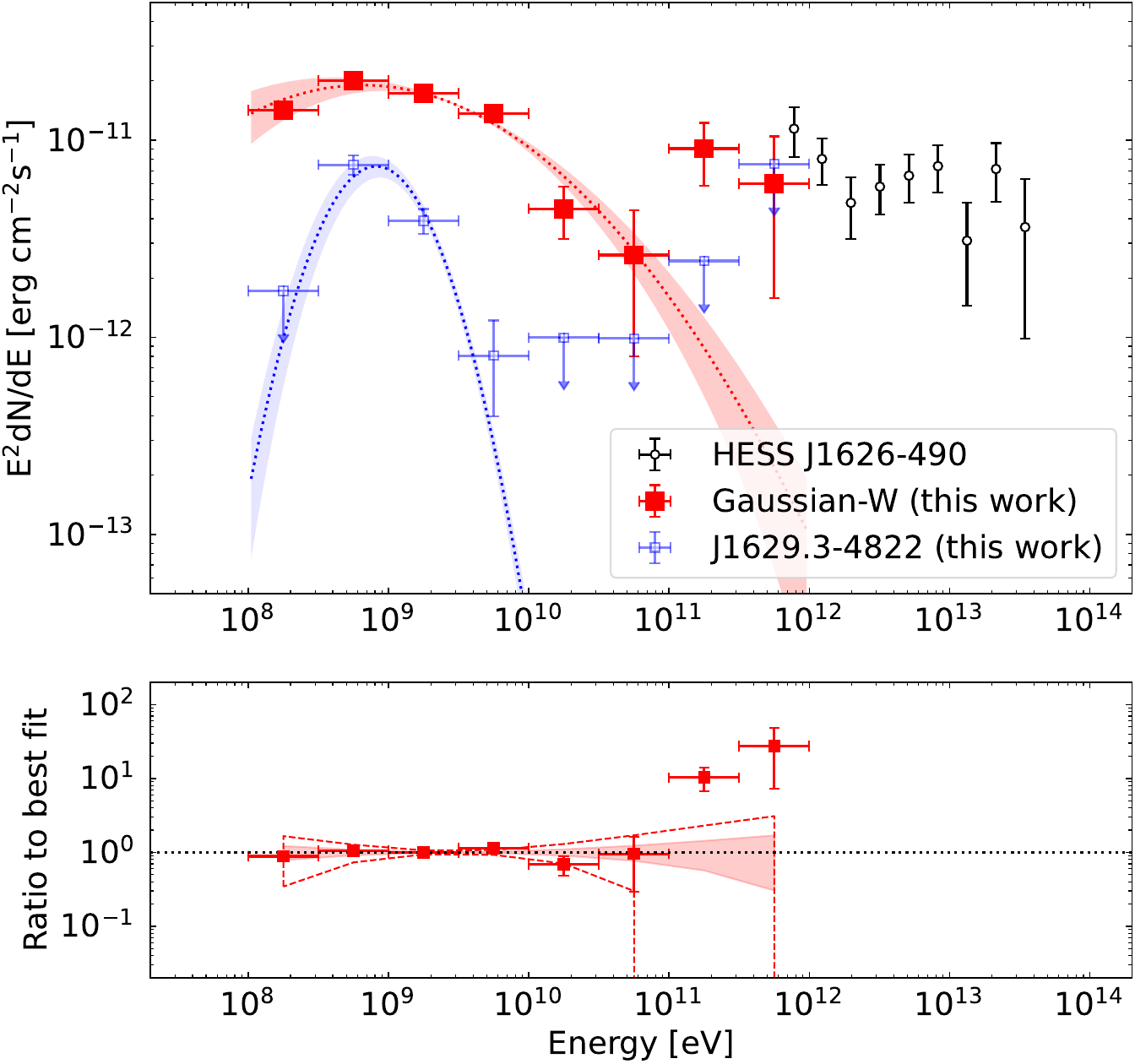}
    \caption{Gamma-ray spectra in the vicinity of SNR G335.2$+$0.1.
    {\bf Top:} The red-filled and blue-open squares show the {\it Fermi}-LAT spectrum of Gaussian-W and J1629.3$-$4822c, respectively. 
    The red-dashed and blue-dotted lines represent the best-fit results using the log-parabola function for each spectrum.
    The open circles show those for HESS J1626$-$490~\citep{HESS2018}.
    {\bf Bottom:} The residual from the fitted function of the Gaussian-W spectrum.
    The meanings of the red markers and the dashed line are the same as in the top panel.
    The shaded region and the area enclosed by the dashed lines are the $1\sigma$ and $3\sigma$ statistical uncertainty of the fit function, respectively.  
    }
    \label{fig:Fermi_Spectrum}
\end{figure}

We extract GeV gamma-ray spectra at the locations of SNR G335.2$+$0.1 and HESS J1626$-$490.
As for the spatial assumption, we adopt the 1-Gaussian+1-PS model for these sources, respectively.
During the likelihood fitting, the spectral parameters of background sources (Galactic diffuse, isotropic background, and 4FGL sources detected in the catalog at $10\sigma$ or higher, are set to free.
Fig.~\ref{fig:Fermi_Spectrum} shows the energy spectra of the two sources as well as the H.E.S.S. data \citep{HESS2007}.
We fit the energy spectrum of Gaussian-W using a log-parabolic function instead of a simple power-law function because the TS value in the log-parabola fit is significantly ($\sim14\sigma$) improved from that with a power-law function.
Meanwhile, the observed spectrum shows a clear discrepancy compared to the fitted function at $> 50~\rm{GeV}$, indicating a spectral hardening. 
The flux at $> 100~\rm{GeV}$, which partially overlaps the range of the H.E.S.S. data, is comparable with that of the H.E.S.S. spectrum.

To verify the robustness of this spectral analysis, we check the stability of the spatial parameters in the 1-Gaussian+1-PS model and evaluate the impact of variable spatial parameters on the spectral analysis.
We originally selected the coordinates of the point source from one of 4FGL point sources, whereas here, we allow it to vary freely.
By rescanning the spatial parameters, the best-fit parameters are obtained as (RA, DEC, $\sigma$) = ($246.94^{\circ} \pm 0.04^{\circ}$, $-48.95^{\circ} \pm 0.02^{\circ}$, $0.28^{\circ} \pm 0.02^{\circ}$) for the Gaussian source and (RA, DEC) = ($247.83^{\circ} \pm 0.03^{\circ}$, $-48.47^{\circ} \pm 0.01^{\circ}$) for the point source.
The Gaussian parameters are consistent with the original values within $2\sigma$ error, while the coordinate of the point source is closer to 4FGL J1631$-$4826c than those of J1629$-$4822c.
Even if we adopt these newly obtained spatial parameters, the energy spectrum of Gaussian-W agrees well with the original one within $1\sigma$.

In addition, we attempt to extract spectra of the SNR and H.E.S.S. source regions individually using the 2-Gaussian model (even though the model is not statistically favored from the Likelihood test).
The details are described in Appendix \ref{section:AppendixA}.
We find a clear difference in spectra between the two sources at energies greater than 10~GeV, and the SNR region shows a lower flux than the H.E.S.S. source in this energy range.
However, in the energy band below 10~GeV, the two sources have similar spectra, indicating a failure to separate the emissions.
We have tried using better PSF class data (PSF2 and PSF3) and a reoptimized spatial model, and they yield similar results.
In summary, at $>10$~GeV, only the emission from the H.E.S.S. source region is significantly detected; therefore, we do not need to consider the contribution of the SNR region in the spectral analysis. 
However, in the lower energy range below 10~GeV, the current limited angular resolution makes it difficult to spatially resolve the emissions from the H.E.S.S. and SNR region.

\subsection{X-ray data analysis} \label{sec:ana:X_ray}

\begin{figure}
    \centering
\includegraphics[width=\hsize]{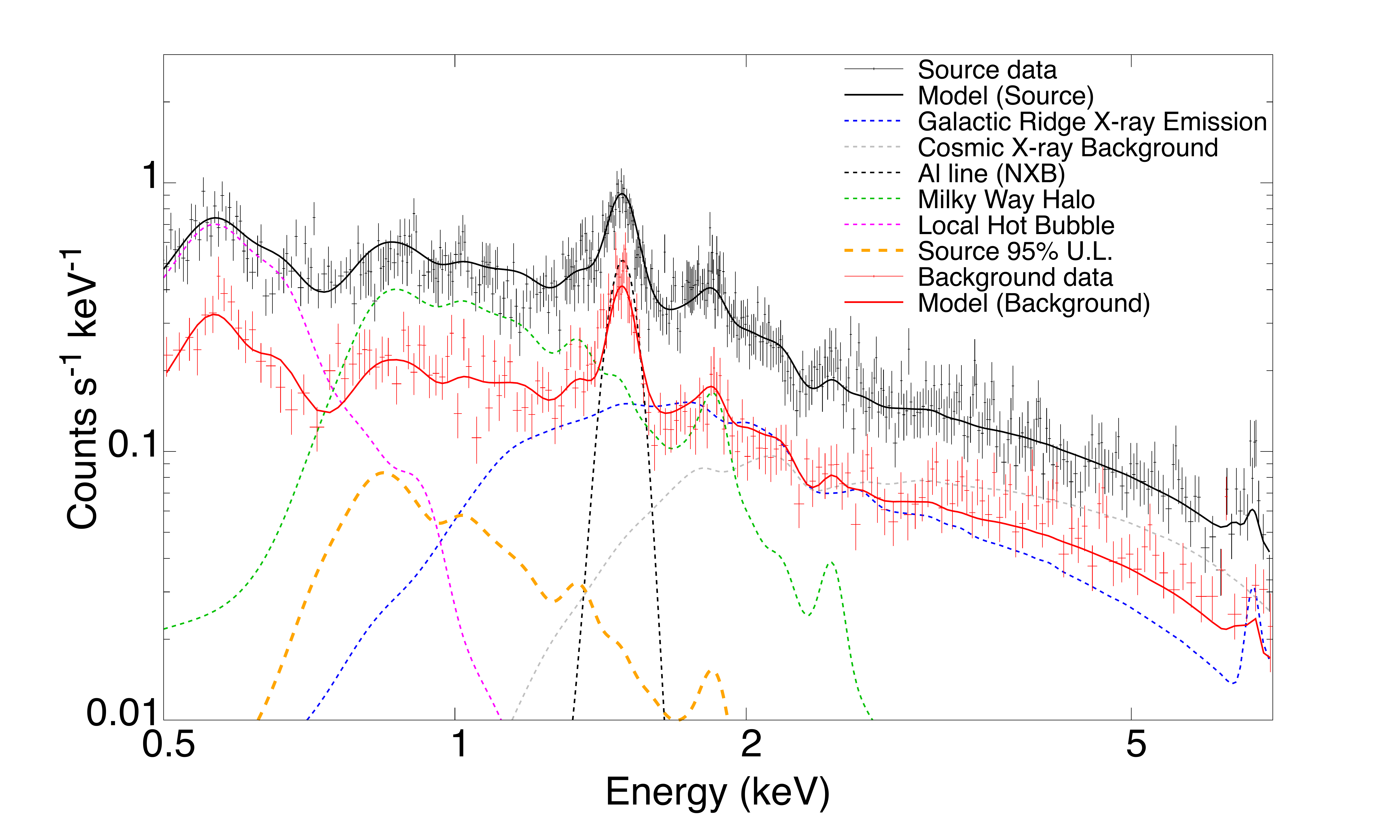}
    \caption{X-ray spectra of the G335.2+0.1 (source; black) and background (red) regions obtained with XMM-Newton and the best-fit models. The solid lines show the total models for the two regions, whereas the dashed lines represent the model components for the source region. The orange dashed line indicates the G335.2+0.1 model with the 95\% upper limit flux.
    }
    \label{fig:X_ray_spectrum}
\end{figure}

We create 0.5--5.0~keV images from the XMM data covering the entire radio shell of G335.2+0.1, finding no excess emission compared to the background.
Thus, we here aim to derive an upper limit of the source emission.
For the spectral analysis, we select a circular region with a radius of $10\arcmin$ from the center of G335.2+0.1 as the source region, and an annulus with inner and outer radii of $10\farcm5$ and $13\farcm0$, respectively, as the background region.
Because the spatially-dependent detector background of the MOS requires a careful treatment \citep{kuntz08}, we only use the pn in the present work.
Part of the southern edge of the observed area, which is affected by stray light, is excluded from our analysis.
As shown in Fig.~\ref{fig:X_ray_spectrum}, the X-ray spectra extracted in the source and background regions are very similar, confirming the absence of the source emission.

We then model the spectra of the source and background regions simultaneously to quantify the upper limit of the source emission.
Our sky background model includes the Local Hot Bubble, Milky Way Halo (Transabsorption Emission; \citealt{kuntz00, masui09}), Cosmic X-ray Background (CXB; \citealt{barcons00, kushino02}), and Galactic Ridge X-ray Emission \citep{uchiyama13, koyama18}, following a methodology for faint diffuse sources close to the Galactic center (e.g., \citealt{suzuki20a}).
Optically-thin thermal plasma with the collisional ionization equilibrium (CIE) is assumed for the Local Hot Bubble ($kT_{\rm e} = 0.2$~keV; without interstellar absorption), Milky Way Halo ($kT_{\rm e} = 0.7$~keV; with a free absorption column $N_{\rm H,MWH}$), and Galactic Ridge X-ray Emission ($kT_{\rm e} = 7$~keV; with the absorption $N_{\rm H,GR} = N_{\rm H,MWH}$). The solar abundance of metals is assumed for all these models. 
CXB is modeled with a power-law with a photon index of 1.4 and absorption column $N_{\rm H,CXB} = 2 \times N_{\rm H,MWH}$.
The surface brightnesses of these components (four in total) are treated as free parameters and are tied between the source and background regions except for CXB, for which both are allowed to vary. This treatment is to explain the slightly higher flux of the background region at the energies above $\sim 3$~keV.
The detector background (NXB) spectra are created with the {\tt pn\_back} tool for both source and background regions, and are considered via the $W$-stat in XSPEC. We add a zero-width Gaussian model to compensate the lack of an Al line at 1.49~keV, which is not included in the output of the {\tt pn\_back} tool. The normalization is allowed to fit the data.

As for the source, we assume an optically-thin, CIE plasma with the solar abundance.
The electron temperature is assumed to be $kT_{\rm e} = 0.2$--$1.0~\rm{keV}$, corresponding to those seen in typical middle-aged to old SNRs.
The interstellar absorption column density is fixed to $1 \times 10^{22}$~cm$^{-2}$, which is a typical value based on an assumption of an interstellar electron density of 1~cm$^{-3}$ and a distance of 3.3~kpc (see Sect.~\ref{sec:ana:radio} for the distance estimate).
The resultant upper limit of the unabsorbed flux of the source is $\approx 1 \times 10^{-12}$~erg s$^{-1}$ cm$^{-2}$ in 0.5--7.0~keV ($95\%$ confidence level).
The background fluxes are confirmed to be roughly consistent with previous studies, i.e., LHB flux of $\sim 1\times10^{-15}$~erg~s$^{-1}$~cm$^{-2}$~arcmin$^{-2}$ in 0.3--1.0~keV \citep{masui09}, Milky Way Halo flux of $\sim 3\times10^{-15}$~erg~s$^{-1}$~cm$^{-2}$~arcmin$^{-2}$ in 0.1--2.0~keV \citep{kuntz00}, CXB flux of $\approx 5.4\times10^{-15}$~erg~s$^{-1}$~cm$^{-2}$~arcmin$^{-2}$ in 2--10~keV \citep{kushino02}, and Galactic Ridge X-ray flux of $\sim 1\times10^{-14}$~erg~s$^{-1}$~cm$^{-2}$~arcmin$^{-2}$ in 5--8~keV at the source location \citep{uchiyama13}.
The flux difference of CXB between the source and background regions is found to be $\approx 20\%$, which is reasonable on consideration of the relatively small areas of our analysis regions.
The plasma density is estimated from the emission measure to be $n_{\rm e} \lesssim 0.3~{\rm cm^{-3}} (d/3.3~{\rm kpc})^{-0.5} (f/1)^{-0.5}$, where $d$ is the distance and $f$ is a volume filling factor of the SNR plasma.
We note that the volume filling factor may be smaller and thus the plasma density may be higher, based on the cases of the remnants with faint thermal emission such as RX~J1713.7$-$3946 \citep{katsuda15}, or relatively young SNRs with thin shells such as RCW~86 \citep{williams11, suzuki22}.

\subsection{Radio data analysis} \label{sec:ana:radio}

\begin{figure*}
    \begin{tabular}{cc}
        \begin{minipage}{0.47\hsize}
            \centering
            \includegraphics[width=\hsize]{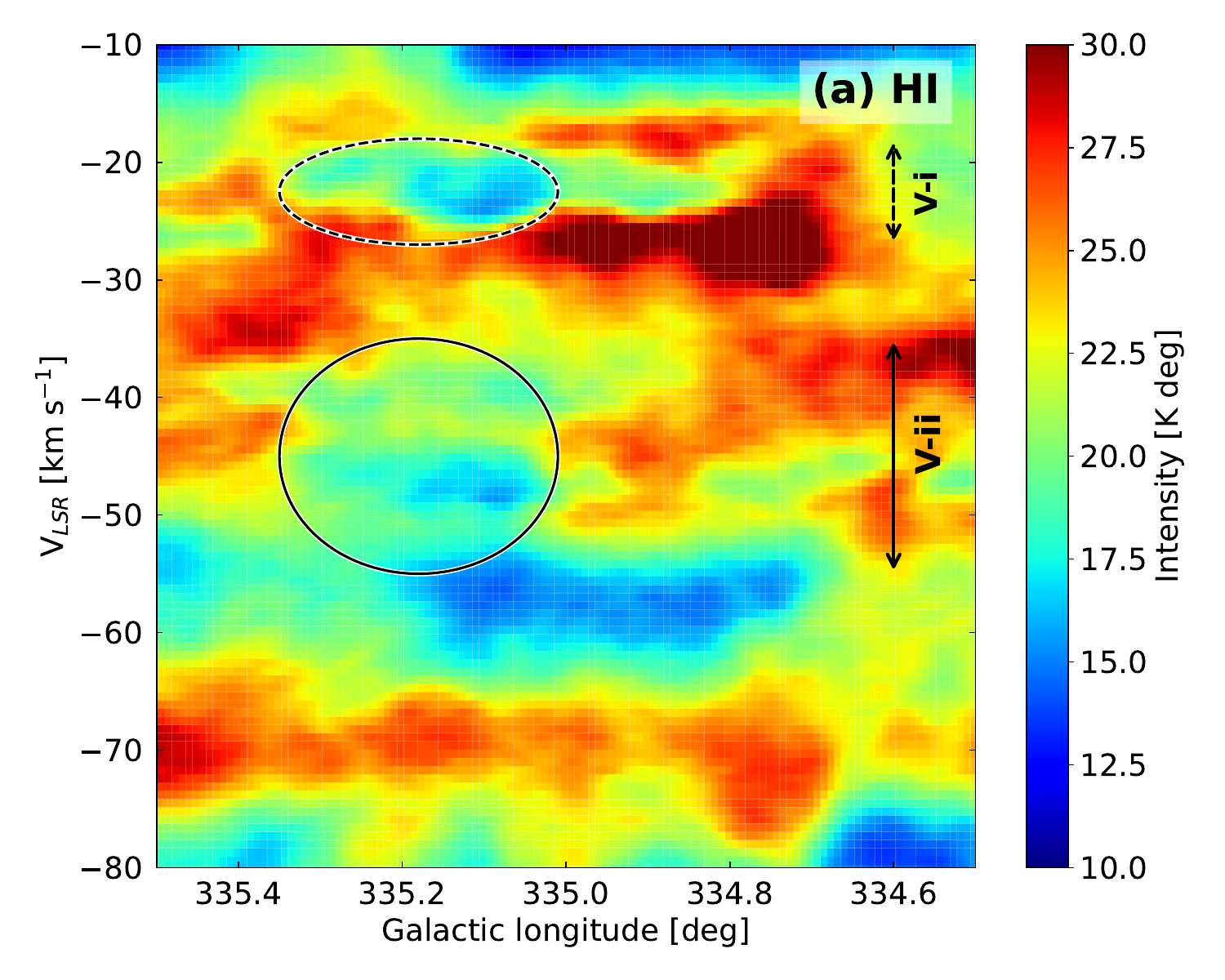}
        \end{minipage}
        &
        \begin{minipage}{0.47\hsize}
            \centering
            \includegraphics[width=\hsize]{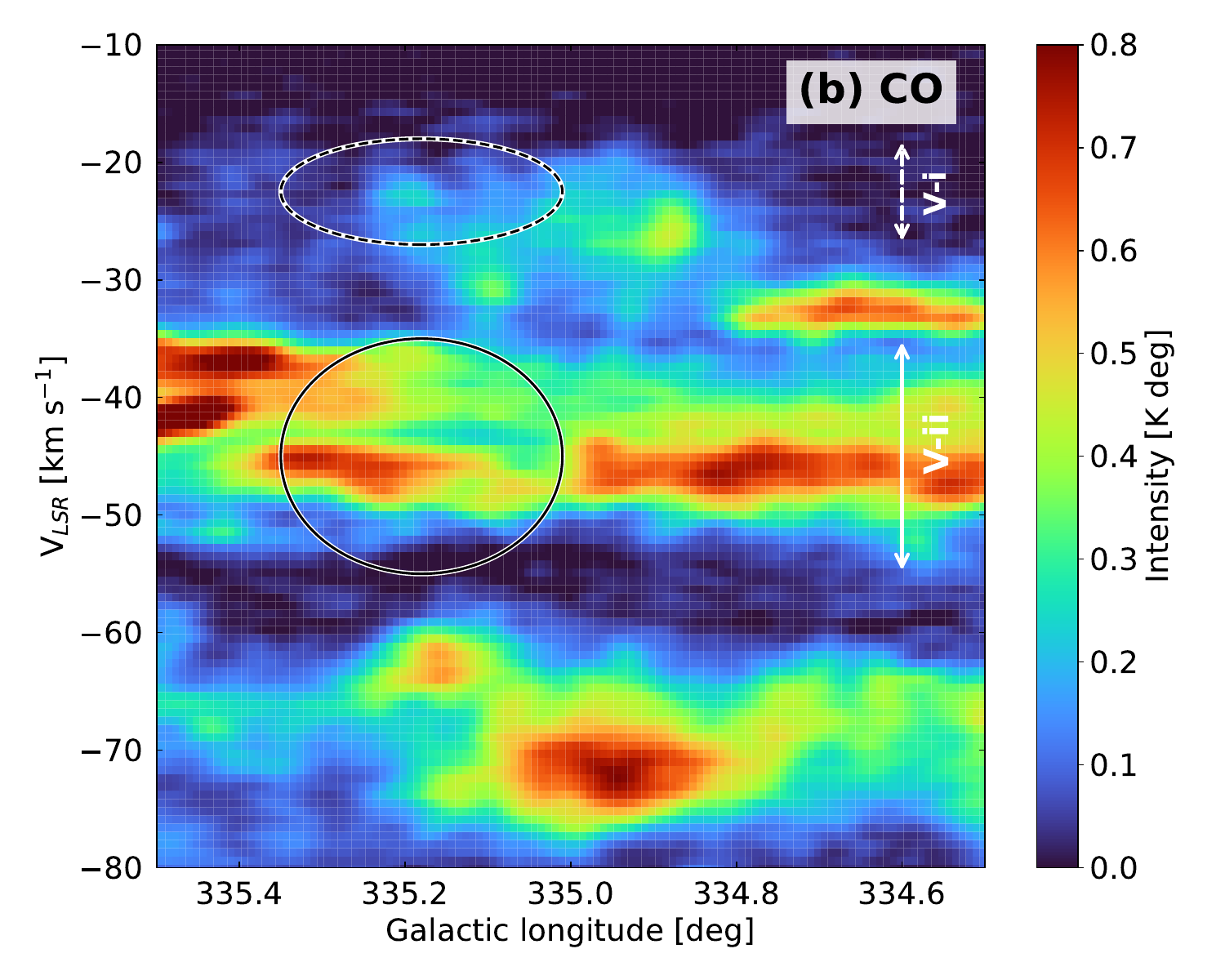}
        \end{minipage}
    \end{tabular}
\caption{Longitude-velocity diagrams of H\,{\sc i} (a) and CO (b).
    The integration range is from 0.0$^{\circ}$ to 0.25$^{\circ}$ in Galactic latitude. 
    The dashed and solid ellipses indicate the boundaries of the H\,{\sc i} cavities, and their range in the x-axis direction correspond to the extent of SNR G335.2+0.1.
    The dashed arrow labeled V-i shows the velocity range ($-27$--$-18~{\rm km\,s^{-1}}$) pointed out by \cite{Eger2011}, while the solid arrow labeled V-ii is the one ($-55$--$-35~{\rm km\,s^{-1}}$) newly found in this work.
}
\label{fig:radio:PVdiagram}
\end{figure*}

\begin{figure*}
    \begin{tabular}{cc}
        \begin{minipage}{0.47\hsize}
            \centering
            \includegraphics[width=\hsize]{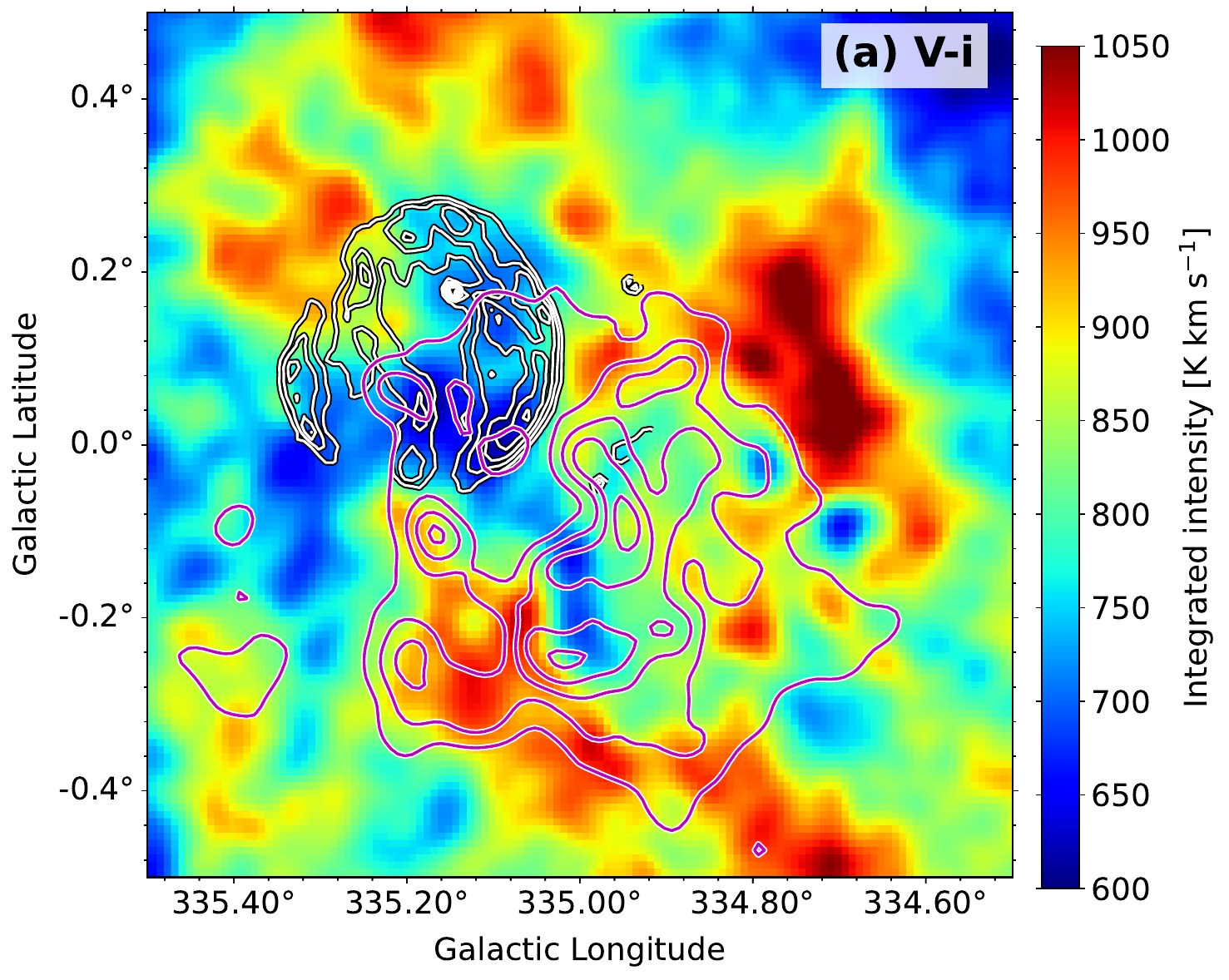}
        \end{minipage}
        &
        \begin{minipage}{0.47\hsize}
            \centering
            \includegraphics[width=\hsize]{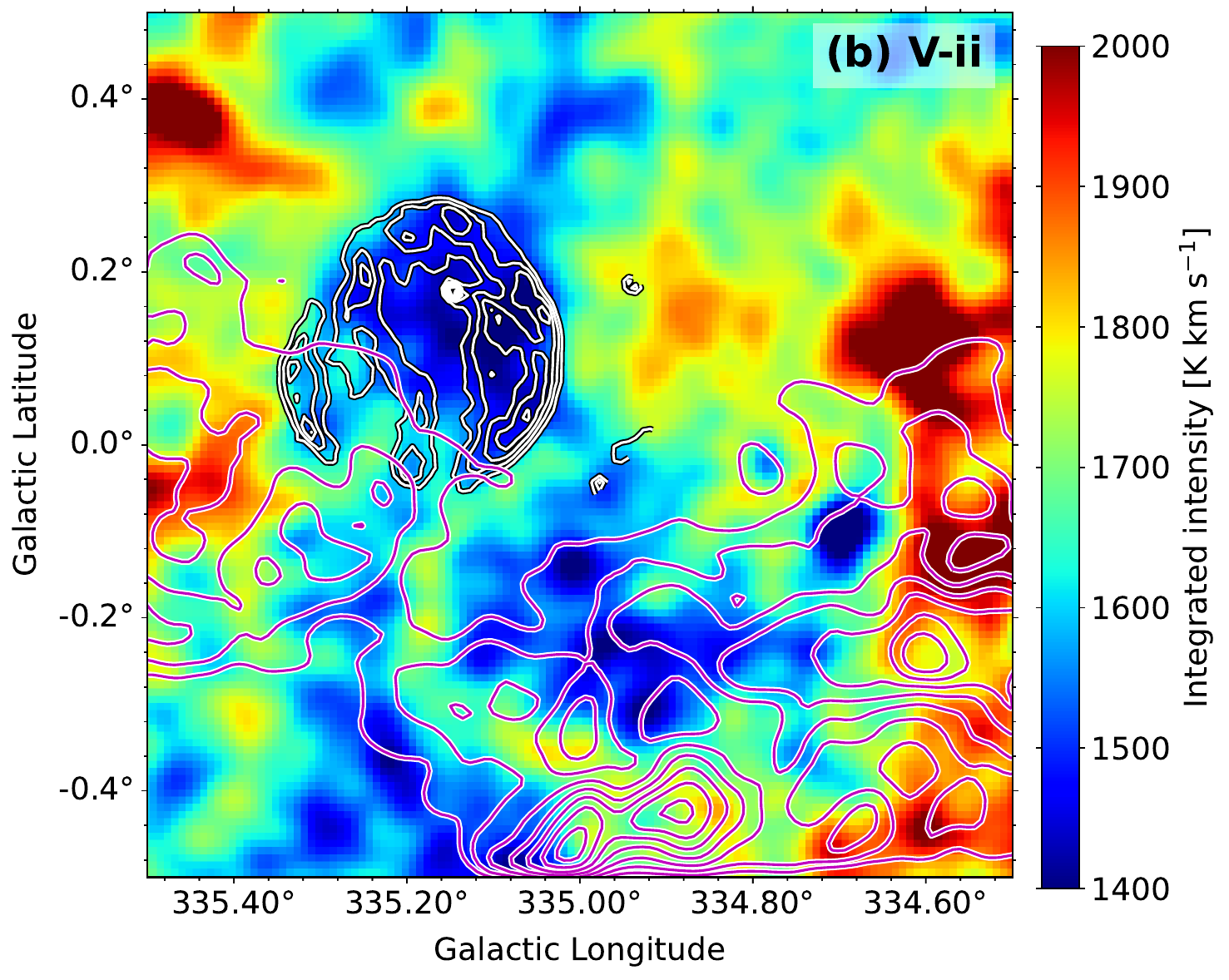}
        \end{minipage}
  \end{tabular}
\caption{Integrated intensity maps of H\,{\sc i} line in the vicinity of SNR G335.2+0.1.
    The integration velocity ranges are $-27$--$-18~{\rm km\,s^{-1}}$ (labeled V-i) and $-55$--$-35~{\rm km\,s^{-1}}$ (labeled V-ii) for panels (a) and (b), respectively.
    The magenta contours show the $^{12}$CO ($J=1-0$) line intensity and are linearly spaced in increments of 5.0~K km\,s$^{-1}$ from 5.0--25.0~K km\,s$^{-1}$.
    The white contours show the radio continuum emission at 843~MHz with the MOST telescope and are linearly spaced in increments of 0.02~K from 0.02--0.12~K.
}
\label{fig:radio:2Dmap}
\end{figure*}

To search for clouds associated with the SNR, we scan the entire local standard of rest velocity ranges ($V_{\rm LSR} = -150$ -- $+50~\rm{km\,s^{-1}}$) of the H\,{\sc i} and CO data.
Fig.~\ref{fig:radio:PVdiagram} shows the position-velocity ($l$-$v_{\rm LSR}$) diagram around SNR G335.2+0.1, integrated over the Galactic latitude range 0.0$^{\circ}$ to 0.25$^{\circ}$.
In the H\,{\sc i} data (Panel a), two cavity-like structures with a similar extent to the SNR shell are seen at velocities of (i) $V_{\rm LSR} {\sim} -22~\rm{km\,s^{-1}}$ and (ii) $-45~\rm{km\,s^{-1}}$, respectively.
The former velocity (i) is consistent with that pointed out by \cite{Eger2011}.
Such a cavity-like structure is partially seen in the CO data at $V_{\rm LSR} = -45~\rm{km\,s^{-1}}$.
Fig.~\ref{fig:radio:2Dmap} shows the H\,{\sc i} images integrated over the two velocity ranges, (i) $V_{\rm LSR} = -27$--$-18~{\rm km\,s^{-1}}$ and (ii) $-55$--$-35~{\rm km\,s^{-1}}$, which we label V-i and V-ii, respectively.
The morphology of the H\,{\sc i} emission is in good agreement with the shell structure in both velocity ranges.

\begin{figure*}
    \centering
    \begin{tabular}{cc}
        \begin{minipage}{0.45\hsize}
            \centering
            \includegraphics[width=\hsize]{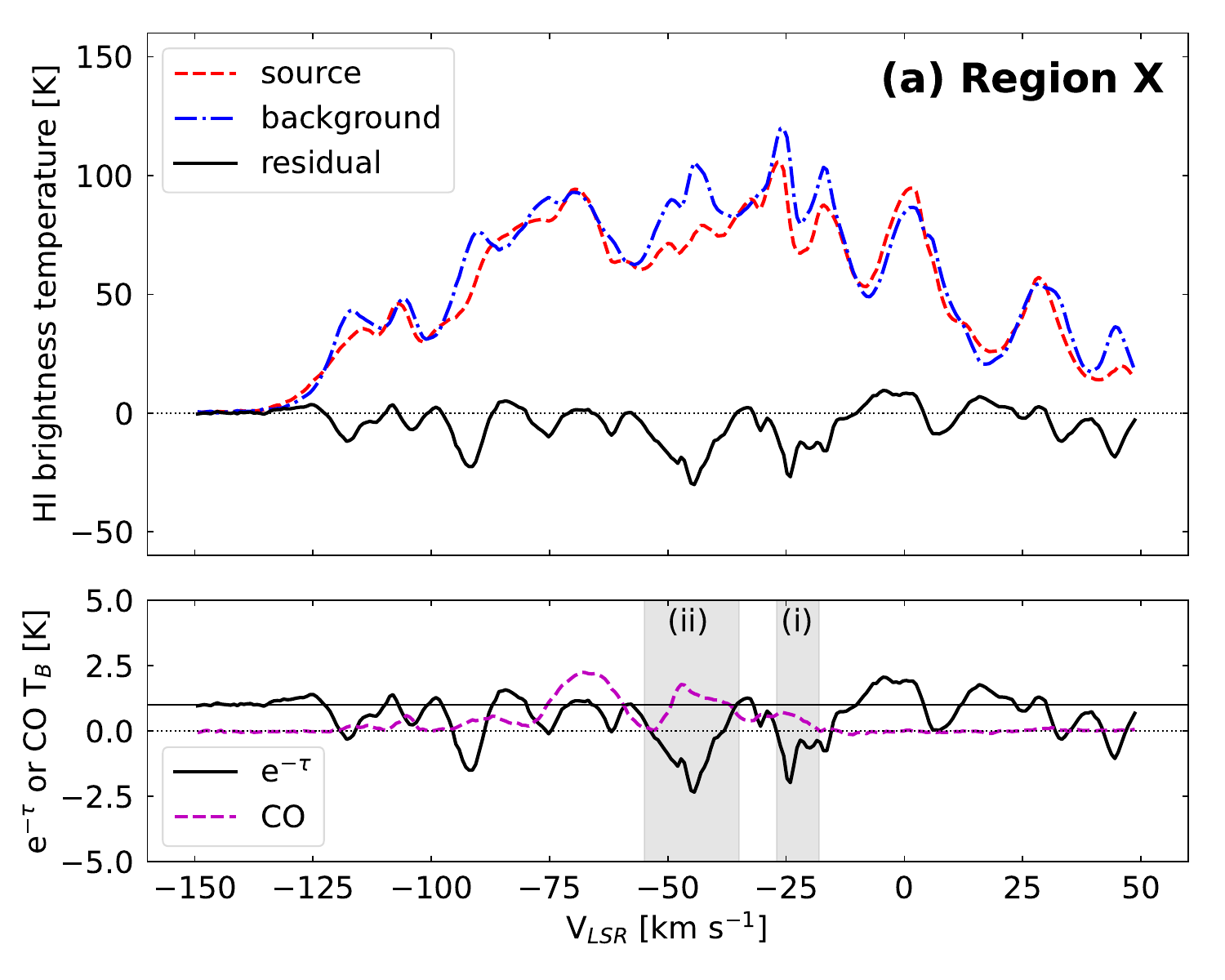}
        \end{minipage}
        &
        \begin{minipage}{0.45\hsize}
            \centering
            \includegraphics[width=\hsize]{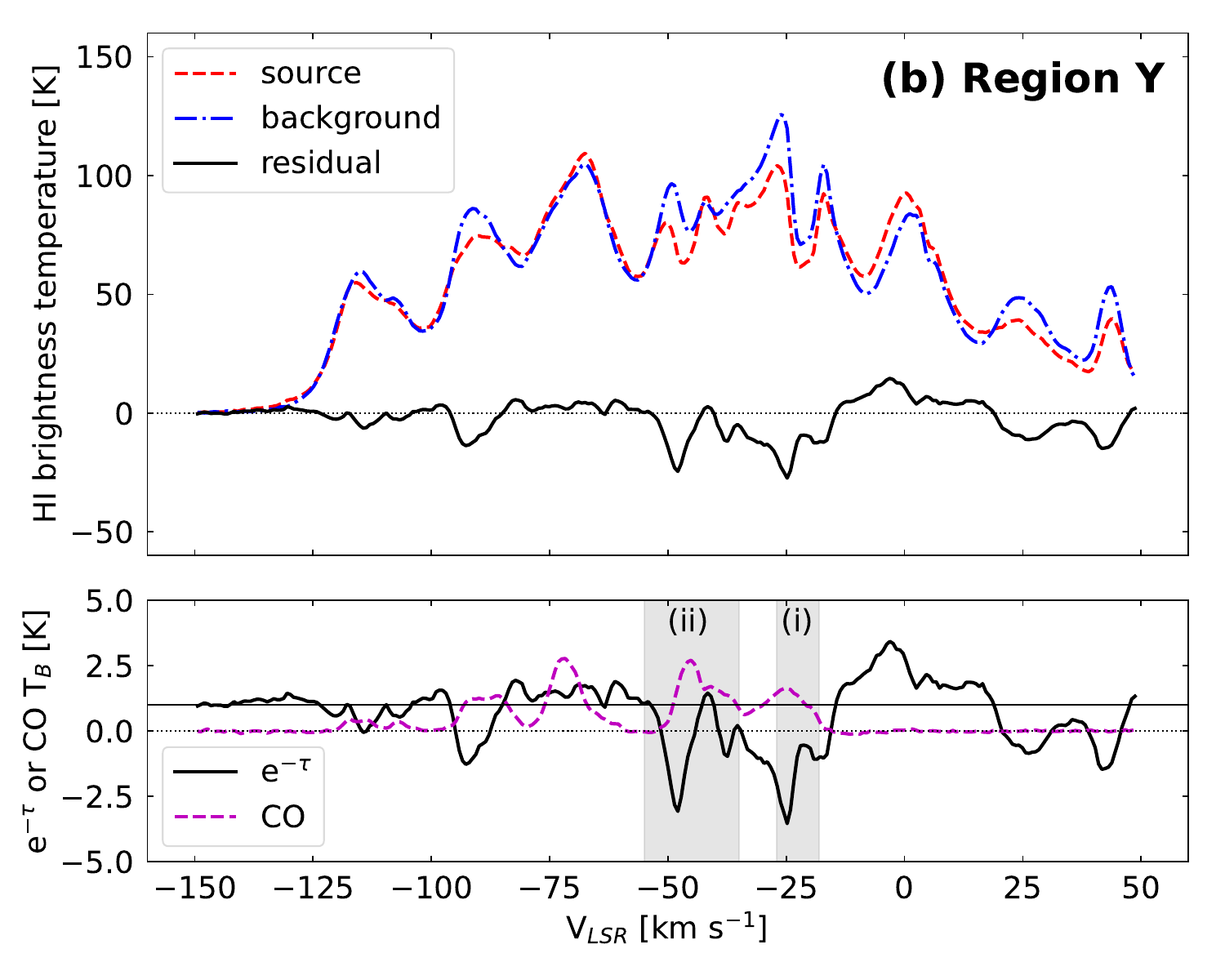}
        \end{minipage}
        \\
        \begin{minipage}{0.45\hsize}
            \centering
            \includegraphics[width=\hsize]{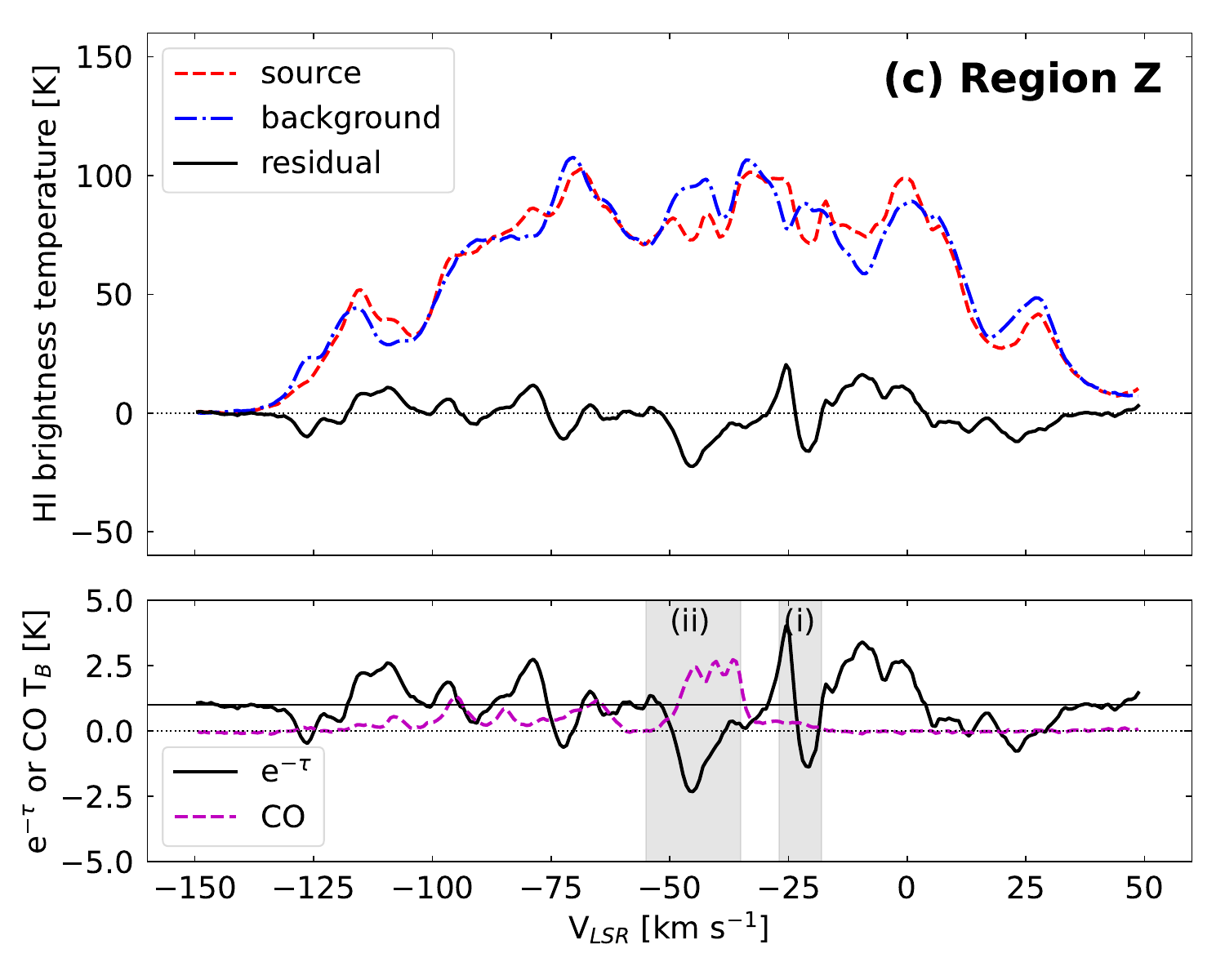}
        \end{minipage}
        &
        \begin{minipage}{0.425\hsize}
            \centering
            \includegraphics[width=\hsize]{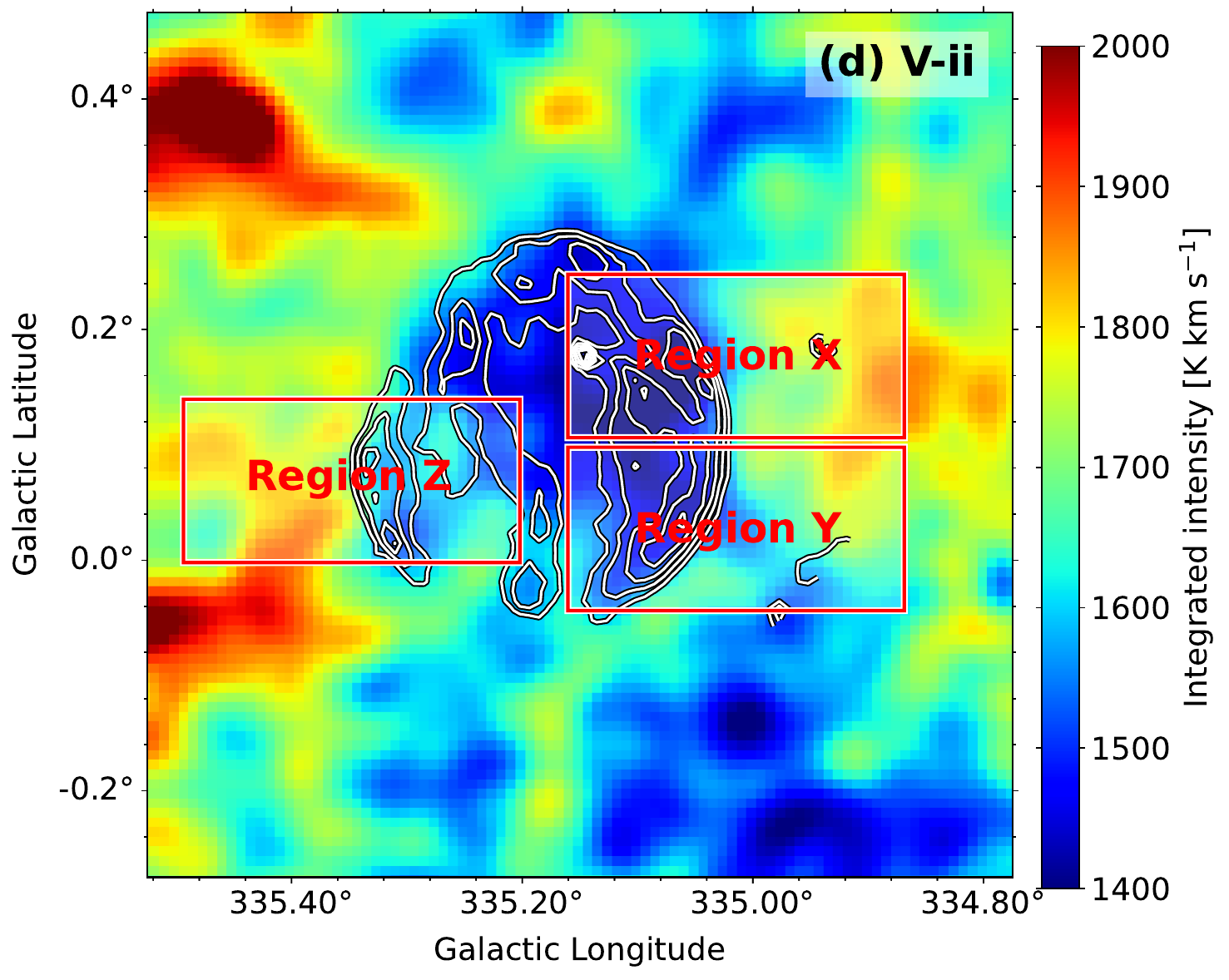}
        \end{minipage}
    \end{tabular}
    \caption{The H\,{\sc i} and CO profiles around SNR G335.2+0.1.
    {\bf (a--c)} The top of each panel shows the H\,{\sc i} profiles of the source region (red dashed), the background region (blue dot-dashed), and the residual (solid black), while the bottom indicates the H\,{\sc i} absorption profiles of $e^{- \tau}$ (black) and the $^{12}$CO ($J=1–0$) line emission (magenta dashed). 
    The gray shaded areas represent velocity ranges that show H\,{\sc i} absorption features in all three regions and correspond to that (i) pointed out by \cite{Eger2011}  and (ii) in this work.
    {\bf (d)} Three rectangular regions X--Z used for the absorption study overlaid on the integrated H\,{\sc i} intensity map. The integration range and the white contours are the same as in Fig.~\ref{fig:radio:2Dmap}.
    }
    \label{fig:radio:absorption}
\end{figure*}

In order to distinguish which velocity band is truly associated with the SNR, we study the H\,{\sc i} absorption line, following \cite{Leahy2010} and \cite{Ranasinghe2017}.
If there is an H\,{\sc i} cloud between the SNR and the observer, the H\,{\sc i} cloud will absorb the continuum emission at 1420~MHz from the SNR.
According to \cite{Ranasinghe2017}, the H\,{\sc i} absorption spectrum is given as:
\begin{equation}
    e^{-\tau} - 1 = \frac{T_{\rm{src}} - T_{\rm{bg}}}{T_{\rm{src}}^{\rm{C}} - T_{\rm{bg}}^{\rm{C}}},
\end{equation}
where $\tau$ is the optical depth of the continuum source, and $T$ ($T^{\rm{C}}$) is the brightness temperature of the line (continuum) emission, while the subscripts, src and bg, indicate the source and background, respectively.
We selected three regions, labeled X, Y, and Z, toward the SNR (Fig.~\ref{fig:radio:absorption} d), in which the inner and outer half are the source region and the background region, respectively.
Panels a--c in Fig.~\ref{fig:radio:absorption} show the H\,{\sc i} absorption spectra of the selected regions.
We find significant absorption features of H\,{\sc i} spectra at two velocity ranges of $V_{\rm LSR} = -27$--$-18~{\rm km\,s^{-1}}$ and $-55$--$-35~{\rm km\,s^{-1}}$ in all the three regions.
These two velocity ranges are identical to those of the HI cavities found in the longitude-velocity diagram.
Fig.~\ref{fig:radio:absorption} also shows the CO spectra and reveals no strong emission in their ranges, supporting the interpretation that the absorption profiles are due to the radio continuum emission from SNR G335.2+0.1 rather than the self-absorption effect due to dense clouds.

Since this SNR is located toward the Galactic center ($l > 270^{\circ}$), each radial velocity corresponds to two kinematic distances.
By adopting the Galactic rotation curve model with the IAU-recommended values of $R_{0} = 8.5$~kpc and $\Theta_{0} = 220$~km\,s$^{-1}$ \citep{Brand1993A&A}, the kinematic distances ($d$) for $V_{\rm LSR} = -45~\rm{km\,s^{-1}}$ are computed to be $d$ = 3.3 and 12.2~kpc as the near side and far side distance, respectively, while for $V_{\rm LSR} = -22~\rm{km\,s^{-1}}$, $d$ = 1.8 and 13.6~kpc.
If the SNR is located at the far side of the Galactic center from us, absorption line features will be detected in the range between $-100~\rm{km\,s^{-1}}$ and $-55~\rm{km\,s^{-1}}$.
However, such a feature cannot be seen in Fig.~\ref{fig:radio:absorption}, suggesting that the SNR is located on the near side.
We thus suggest that the H\,{\sc i} cavity-like structure at $-45~\rm{km\,s^{-1}}$ is associated with the SNR, and the one at $-22~\rm{km\,s^{-1}}$ is caused by the absorption from the continuum emission.
We also evaluate the uncertainty in estimating the kinematic distance, assuming that the minimum and maximum velocities of the cavity are the errors in the systemic velocity.
The resultant systemic velocity and the kinematic distance of SNR G335.2+0.1 are $-45 \pm 10~\rm{km\,s^{-1}}$ and $3.3 \pm 0.6$~kpc, respectively.
The latter is consistent within $2\sigma$ uncertainties with the estimates in the previous studies \citep{Pavlovic2013, Wang2020} and that of PSR J1627$-$4845 \citep{Kaspi1996}. 

Now that we have obtained the distance to the object, we can determine the size and give a rough estimate of the SNR age ($t_{\rm age}$).
Using the kinematic distance, the SNR radius is estimated as $R = 9.6 \pm 1.8$~pc.
The empirical age-size relation of Galactic SNRs \citep[Eq.~2 in][]{Ranasinghe2023ApJS} suggests that the age of an SNR with a radius of 9.6~pc is $t_{\rm age} \sim 4.9$~kyr.
More generally, $t_{\rm age}$ is a function of a SN explosion energy ($E_{\rm SN}$), the hydrogen density within the stellar wind bubble ($n$), as well as the SNR radius.
The plasma density in the bubble is $n \lesssim 0.3~\rm{cm^{-3}}$, as derived in Sect.~\ref{sec:ana:X_ray}.
Once we assume that $E_{\rm SN} = 3 \times 10^{50}~\rm{erg}$ and this SNR is in the Sedov stage \citep{Sedov1959book}, we obtain a dynamical age similar to the previous estimate: 
\begin{equation} \label{eq:t_age}
     t_{\rm age} \sim 5~{\rm kyr} \left( \frac{E_{\rm SN}}{3 \times 10^{50}~{\rm erg}} \right)^{-0.5} \left( \frac{n}{0.3~{\rm cm^{-3}}} \right)^{0.5} \left( \frac{R}{9.6~{\rm pc}} \right)^{2.5}.
\end{equation}
Note that the characteristic age of PSR J1627-4845 (2.7~Myr) is different from the estimated SNR age ($\sim$ 5~kyr); this discrepancy may be due to an overestimation of the characteristic age \citep{Kaspi1996}.

\begin{figure}
    \centering
    \includegraphics[width=\hsize]{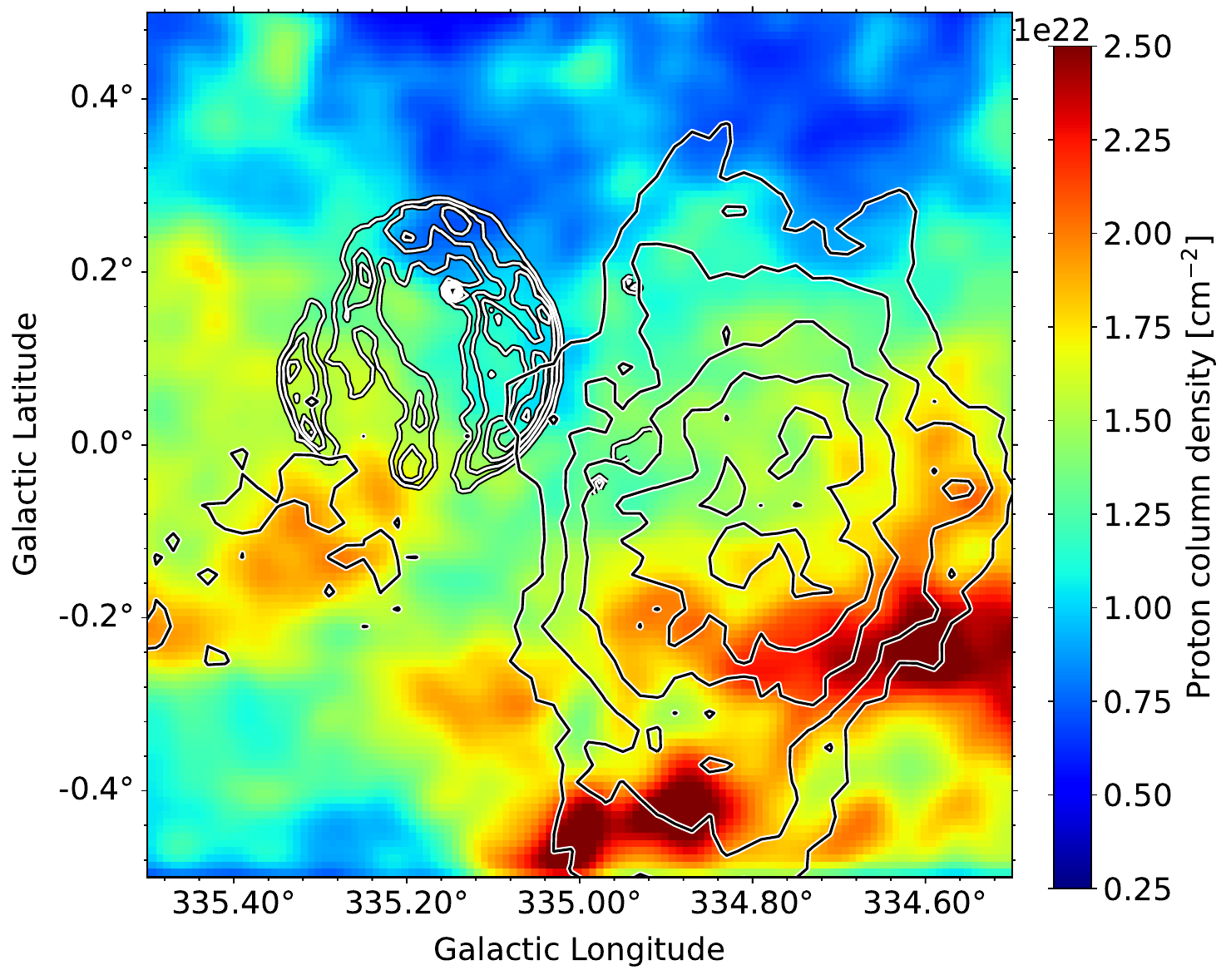}
    \caption{Proton column density map around SNR G335.2+0.1. 
    The white contours indicate the radio continuum emission at 843~MHz as in Fig.~\ref{fig:radio:2Dmap}, while the black contours show the significance level of TeV gamma-ray emission obtained by H.E.S.S. \citep{HESS2018} and are linearly spaced in increments of $1\sigma$ from 3--7$\sigma$.}
    \label{fig:radio:protondensity}
\end{figure}

We calculate the gas density around HESS J1626$-$490.
As for the conversion factor from the integrated intensity to column density, we adopt $X_{\rm CO} = 2 \times 10^{20}~\rm{cm^{-2}\,(K\,km\,s^{-1})^{-1}}$ for $^{12}$CO ($J=1-0$) line \citep{Bertsch1993ApJ} and $X_{\rm HI} = 1.823 \times 10^{18}~\rm{cm^{-2}\,(K\,km\,s^{-1})^{-1}}$ for the H\,{\sc i} line \citep{Dickey1990ARA&A}, respectively, assuming that these emission lines are optically thin.
Fig.~\ref{fig:radio:protondensity} shows the proton column density map in the velocity range of $V_{\rm LSR} = -55$--$-35~{\rm km\,s^{-1}}$.
The morphology of the gas distribution is roughly consistent with the TeV emission, indicating that the possible target protons for the hadronic gamma-ray emission process are present in the H.E.S.S. source region.
Assuming that the cloud is a spherical region with the same extent as the H.E.S.S. source, i.e., a radius of 3.3~kpc $\times$ tan(0.2$^{\circ}$) $\sim$ 11.5~pc, we obtain $n_{\rm cl} \sim 510~{\rm cm^{-3}}$ as the proton density in the H.E.S.S. source region.
In the same manner, the gas density in the SNR region is estimated to be $n_{\rm SNR} \sim 320~{\rm cm^{-3}}$.

\section{Spectral modeling} \label{section:modeling}

Our \textit{Fermi}-LAT analysis reveals the extended emission, Gaussian-W, around HESS J1626$-$490.
The energy spectrum of Gaussian-W shows a spectral hardening at 50~GeV and smoothly connects to the HESS spectrum, suggesting the contribution from two components (e.g., the SNR and the molecular cloud).
Furthermore, the emission region of Gaussian-W transitions with energy from near the SNR to the H.E.S.S. source region (Fig.~\ref{fig:GeVmap} and \ref{fig:Fermi_GaussFit}).
Therefore, we can assume that the gamma-ray emission above 50~GeV is mainly from the H.E.S.S. source, whereas the band below is dominated by the emission from the SNR.
Note that there are no other notable candidates for the emission, as mentioned in Sect. \ref{section:Introduction}.
In this work, we explore the possibility that the H.E.S.S. source originates in the cloud illuminated by escaped CRs from SNR G335.2+0.1, by modeling the observed spectrum.
The calculation method of the model spectrum is described in Sect.~\ref{section:modeling:setup}, and its application is described in Sect.~\ref{sec:model_fit}.

\subsection{Model setup} \label{section:modeling:setup}

The method to estimate the gamma-ray emissions from the cloud and the SNR itself is the same as in the previous work \citep{Oka2022}, which is based on \cite{Gabici2009} and \cite{Ohira2010_MNRAS}, and is concisely described in Appendix~\ref{sec:appendix:cr_escape_model}.
Possible mechanisms for non-thermal radiation are electron bremsstrahlung, inverse Compton (IC) scattering, synchrotron for electrons, and $\pi^{0}$-decay for protons.
For the radiation on the cloud, we calculate not only the contribution of the escaped parent electrons, but also synchrotron radiation from secondary electrons produced in p-p collisions (See Appendix~\ref{sec:appendix:secondary_SED}).
We use the radiative code \textit{naima} \citep{Zabalza2015ICRC} to calculate photon spectra.
As for seed photon fields in the IC process, we consider the cosmic microwave background and Galactic far-infrared (FIR) radiation.
We estimate the FIR density using the interstellar radiation field model in the GALPROP package\footnote{\url{https://galprop.stanford.edu/}}.
It contains two models named R12 and F98 \citep{Porter2008, Porter2017ApJ}.
After confirming that the two models give similar results around G335.2+0.1, we use the average of the two models, 0.48~eV cm$^{-3}$ at 45~K, in this paper.

The target gas density in the cloud region is adopted from the radio analysis in Sect. \ref{sec:ana:radio} and is set to $n_{\rm cl} = 510~\rm{cm}^{-3}$, while we treat the density in the SNR region ($n_{\rm SNR}$) as a free parameter even though the estimates from radio and X-ray analyses are $320~\rm{cm}^{-3}$ (Sect.~\ref{sec:ana:radio}) and $0.3~\rm{cm}^{-3}$ (Sect.~\ref{sec:ana:X_ray}), respectively.
The discrepancy between these estimates implies a non-uniform gas distribution around the SNR, and the estimate from the X-ray (radio) analysis represents the density inside (outside) the SNR shock.
We therefore expect that the effective density in the emission region is between the two values.
Note that similar cases have been discussed in previous studies for other SNRs, and the so-called volume filling factor, the ratio of effective density to an estimate from radio analyses, is estimated to be ${\sim} 10\%$ \citep{Uchiyama2010ApJ, MAGIC2012A&A}.

\subsection{Model application} \label{sec:model_fit}

For the electron-to-proton flux ratio (see Eq.~\ref{eq:qs_e} for definition), we consider two cases: $K_{\rm ep} = 0.01$ for the hadron-dominated model and $K_{\rm ep} = 1$ for the lepton-dominated model.
In the former case, the given $K_{\rm ep}$ is consistent with the local CR abundance \citep{Aguilar2016PhRvL}, and hadronic emissions are dominant in the gamma-ray band.
The latter is an extreme assumption, but it is set to test a case in which leptonic emissions are dominant.
The radio data for the shell region are taken from \cite{Clark1975}, \cite{Whiteoak1996}, \citet{Kaspi1996}, and \cite{Green2019JApA}.
The X-ray spectrum for the SNR region is derived in Sect.~\ref{sec:ana:X_ray}, while for the cloud region we adopt the result in \cite{Eger2011}.

Table \ref{tab:modelpara} tabulates the model parameters.
Most of the parameters are adopted based on the observational and theoretical requirements (see Sect.~\ref{sec:ana:radio} and Appendix \ref{sec:appendix:cr_escape_model}), but some parameters are allowed to vary freely.
We determined those fitting parameters in two steps: (1) scan the values with a coarse grid, and then (2) determine the final values with a finer grid. In the first step, among the fitting parameters, we first find the acceleration efficiency ($\eta$) such that the model reproduces the TeV gamma-ray flux.
We then determine the effective gas density ($n_{\rm SNR}$) and magnetic-field strength ($B_{\rm SNR}$) by comparing the fluxes in the radio and GeV bands.
The current maximum energy $E_{\rm now}$ is determined so that the model reproduces the gamma-ray spectral break at ${\sim}10$~GeV.
When determining the final values in the second step, we scanned each parameter with a step size of 1/4 of the original to find the value that gave the smallest residual between the observational data and the model.
The rest of the fitting parameters ($D_{0}$ and $E_{\max}$) affect the spectrum of the cloud region, and their uncertainties will be discussed later.

\begin{figure}
    \begin{tabular}{l}
        \begin{minipage}{0.90\hsize}
            \hspace*{-0.5cm}\includegraphics[width=\hsize]{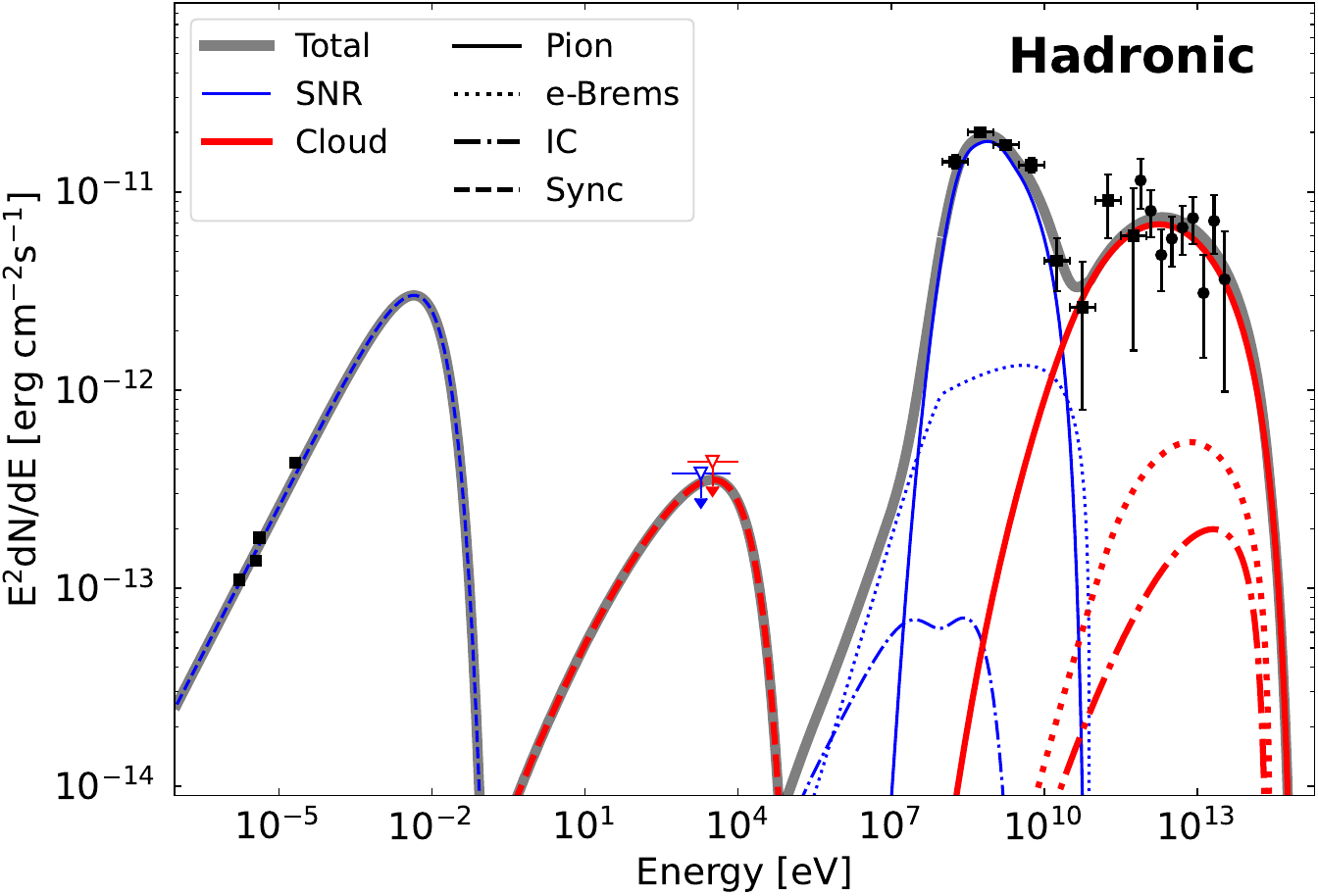}
        \end{minipage}
        \\
        \begin{minipage}{0.90\hsize}
            \hspace*{-0.5cm}\includegraphics[width=\hsize]{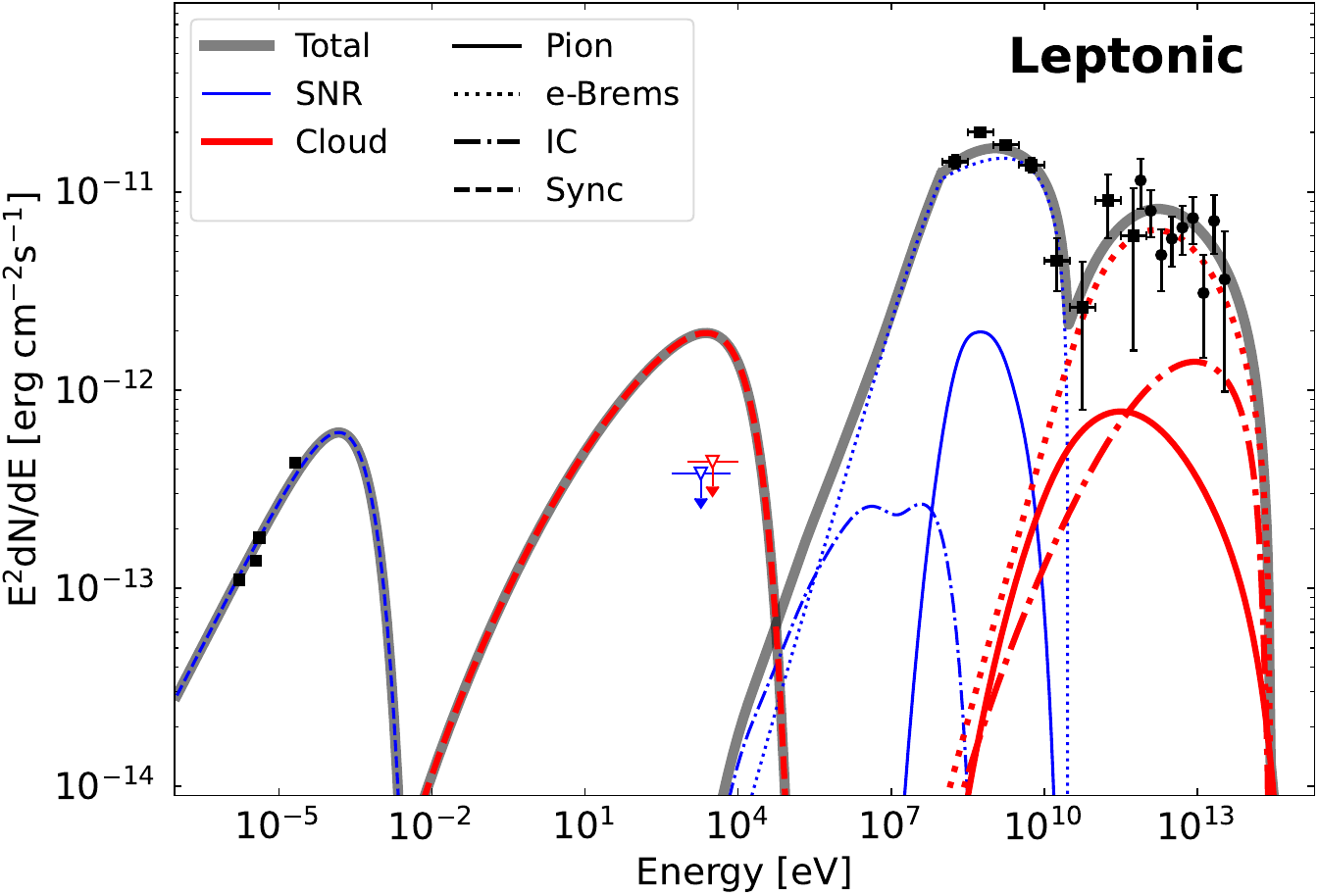}
        \end{minipage}
    \end{tabular}
    \caption{Modeling results for the energy spectrum of Gaussian-W.
    The top panel shows the case of the hadronic-dominated model, while the bottom panel shows the case of the leptonic-dominant case.
    The dashed, dot-dashed, dotted, and solid curves represent the synchrotron, IC, electron bremsstrahlung, and $\pi^{0}$-decay emissions, respectively.
    The blue, red, and black curves show the non-thermal emissions from the shell and cloud and their sum, respectively.
    The data points in the radio band are for the SNR region and taken from the literature \citep{Clark1975, Whiteoak1996, Kaspi1996, Green2019JApA}.
    In the X-ray band, the red upper limit represents the result for the cloud region, which is taken from \cite{Eger2011}, while the blue upper limit is the one for the SNR region, obtained in this work.
    The GeV spectrum is the one of Gaussian-W, while the TeV spectrum is adopted from \cite{HESS2007} as in Fig. \ref{fig:Fermi_Spectrum}.
    \label{fig:modeling:best}
    }
\end{figure}

Fig. \ref{fig:modeling:best} shows the modeling results in both the hadronic-dominated and leptonic-dominated cases, and the values of model parameters are shown in Table \ref{tab:modelpara}.
In the hadronic-dominated case, hadronic emissions can reproduce both gamma-ray spectra at the shell and cloud ($\chi^{2}/{\rm ndf} = 23.4/17$), while in the leptonic-dominated case, electron bremsstrahlung emissions reproduce them ($\chi^{2}/{\rm ndf} = 32.5/17$).
However, the leptonic-dominated model is in tension with the observed upper limit in X-ray, once we assume $B_{\rm cl} \gtrsim 3~{\rm \mu G}$.
Note that the spectral break in the electron-bremsstrahlung spectrum is due to the low-energy threshold for electron energy in clouds.
Since the cosmic ray flux in clouds is reduced by ionization loss below 100~MeV \citep{Phan2018MNRAS}, we adopted a conservative threshold of 100~MeV.
A lower threshold results in a higher MeV gamma-ray flux, but this change does not affect the conclusions of this study.

\begin{table*}  
    \caption{Fiducial parameters used to calculate the CR-escape model spectra. The values in parentheses are used for the leptonic-dominated model; otherwise, the hadronic and leptonic-dominated models use the same values. 
    The fourth column indicates whether each parameter is ``adopted'' from theoretical and observational requirements or ``scanned'' to match to the spectral data.} 
    \label{tab:modelpara}
    \centering
    \begin{tabular}{lccc}
    \hline
    SNR parameters & Symbol & Value & \\\hline
    SN explosion energy & $E_{\rm SN}$ & $0.3 \times 10^{51}~{\rm erg}$ & adopted \\ 
    Initial shock velocity & $u_{\rm 0, sh}$ & $10^{9}~{\rm cm}~{\rm s}^{-1}$ & adopted  \\ 
    Age of the SNR & $t_{\rm age}$ & $5.0\times10^{3}~{\rm yr}$ & adopted   \\ 
    Distance to the SNR & $d$ & $3.3~{\rm kpc}$ & adopted \\ 
    Acceleration efficiency & $\eta$ & 0.1 (0.01) & scanned \\ 
    Electron-to-proton flux ratio & $K_{\rm ep}$ & 0.01 (1) & adopted \\
    Maximum CR energy at $t_{\rm Sedov}$ & $E_{\max}$ & $1 \times 10^{15}~{\rm eV}$ & scanned \\ 
    Maximum CR energy at $t_{\rm now}$ & $E_{\rm now}$ & $8 \times 10^{10}~{\rm eV}$ ($3 \times 10^{10}~{\rm eV}$) & scanned \\ 
    Magnetic field in the SNR & $B_{\rm SNR}$ & $30~{\rm \mu G}$ ($7~{\rm \mu G}$) & scanned \\
    Particle index in the SNR & $p_{\rm SNR}$ & 2.0 & adopted \\
    Particle index after escaping & $p_{\rm esc}$ & 2.3 & adopted \\
    Time when electrons start escaping (normalized to the Sedov age) & $\xi_{\rm e, inj}$ & 1 & adopted \\
    Number density in the SNR region & $n_{\rm SNR}$ & $12~{\rm cm}^{-3}$ ($15~{\rm cm}^{-3}$) & scanned \\ 
    Diffusion coefficient at $10~{\rm GeV}$ & $D_0$ & $2 \times 10^{26}~{\rm cm}^2~{\rm s}^{-1}$ ($4 \times 10^{26}~{\rm cm}^2~{\rm s}^{-1}$) & scanned \\ 
    Slope of diffusion coefficient & $\delta$ & $1/3$ & adopted \\ 
    Magnetic field in ISM & $B_{\rm ISM}$ & $3~{\rm \mu G}$ & adopted \\ 
    Distance to Cloud from SNR & $d_{\rm cl}$ & $23~{\rm pc}$ & adopted \\
    Radius of the Cloud & $R_{\rm cl}$ & $11.5~{\rm pc}$ & adopted \\
    Average hydrogen density of the Cloud & $n_{\rm H}$ & $510~{\rm cm}^{-3}$ & adopted \\
    Magnetic field in Cloud & $B_{\rm cl}$ & $3~{\rm \mu G}$ & adopted \\ 
    \hline \\
    \end{tabular}
\end{table*}

\section{Discussion} \label{section:Discussion}

\subsection{Leptonic vs Hadronic?}

\begin{figure}
    \centering
    \includegraphics[width=0.95\hsize]{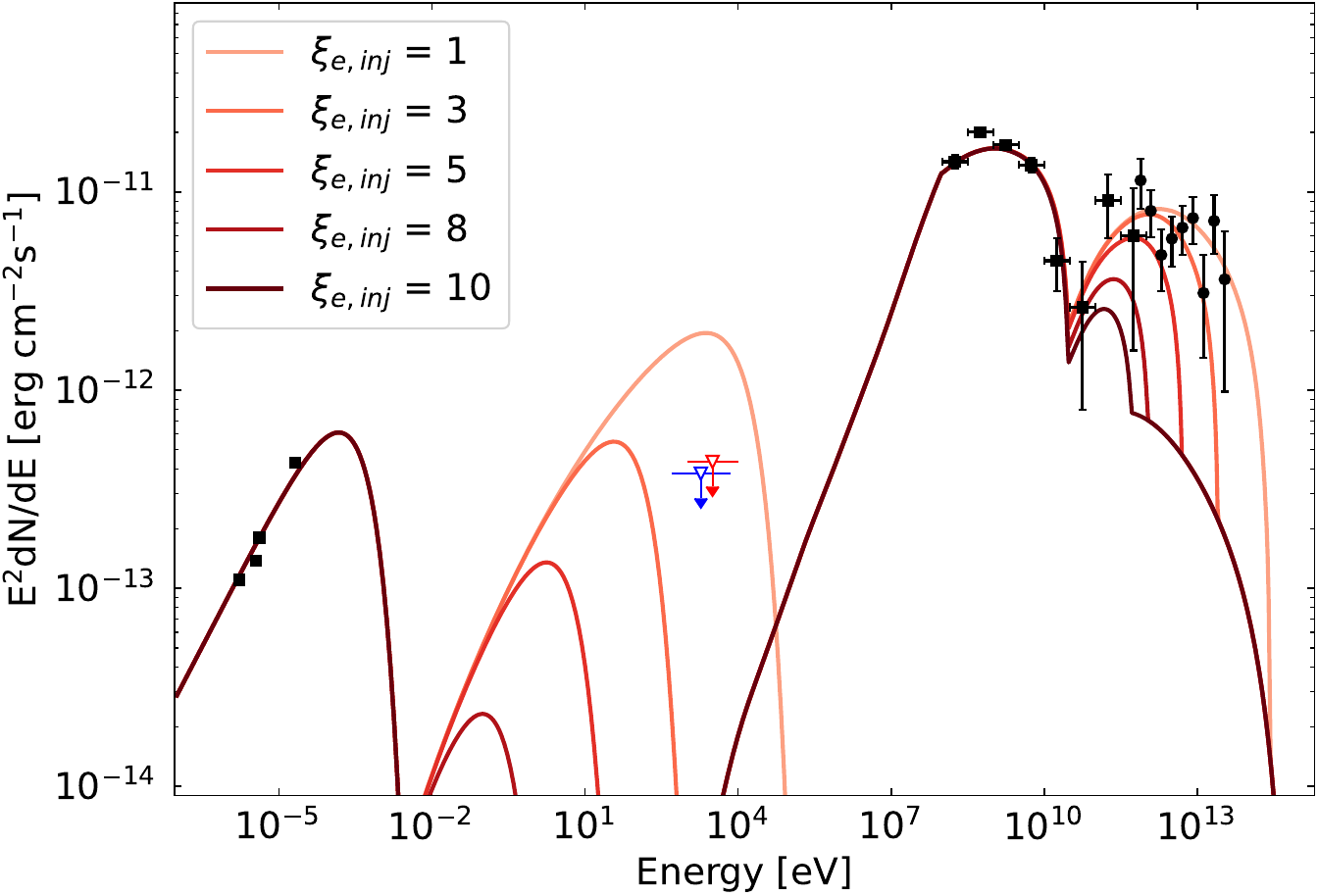}
\caption{The $\xi_{\rm e, inj}$ (scaling factor between the age at which electrons escape and the age at which protons escape, see Appendix B and eq.~\ref{eq:xi_e_inj}) dependency of the model spectrum.
    The data points are the same as in Fig.~\ref{fig:modeling:best}.
    The values used for each parameter are written in the legend of the figure.
}
\label{fig:modeling:leptonic_modeldep}
\end{figure}

First, the tension between the upper limit and the model in the X-ray band is one reason to prefer the hadronic model over the leptonic model.
A further concern in the leptonic-dominated model is the unrealistic assumption of the time when electrons escape from the SNR ($t_{\rm e} = \xi_{\rm e, inj} \times t_{\rm Sedov}$).
When $\xi_{\rm e, inj}$ is set to be large, the escape of high-energy electrons is suppressed.
Fig.~\ref{fig:modeling:leptonic_modeldep} shows the model variations with $\xi_{\rm e, inj}$ for the leptonic-dominated case, fixing the other parameters at the values in Table \ref{tab:modelpara}.
According to \cite{Ohira2012MNRAS}, $\xi_{\rm e, inj}$ depends on the environment of the magnetic field and is expected to be 5--500 under the typical time evolution of the magnetic field, while our model requires $\xi_{\rm e, inj} < 5$ to reproduce the observed spectrum.
Thus, the leptonic-dominated model is unfavored but not ruled out since there are no observational constraints on the special environment of the magnetic field.
This discussion can be simplified if we introduce a variable that links the time evolution of the magnetic field to the maximum acceleration energy, but this is beyond the scope of this paper.

\begin{figure}
    \begin{tabular}{l}
        \begin{minipage}{0.95\hsize}
            \hspace*{-0.5cm}\includegraphics[width=\hsize]{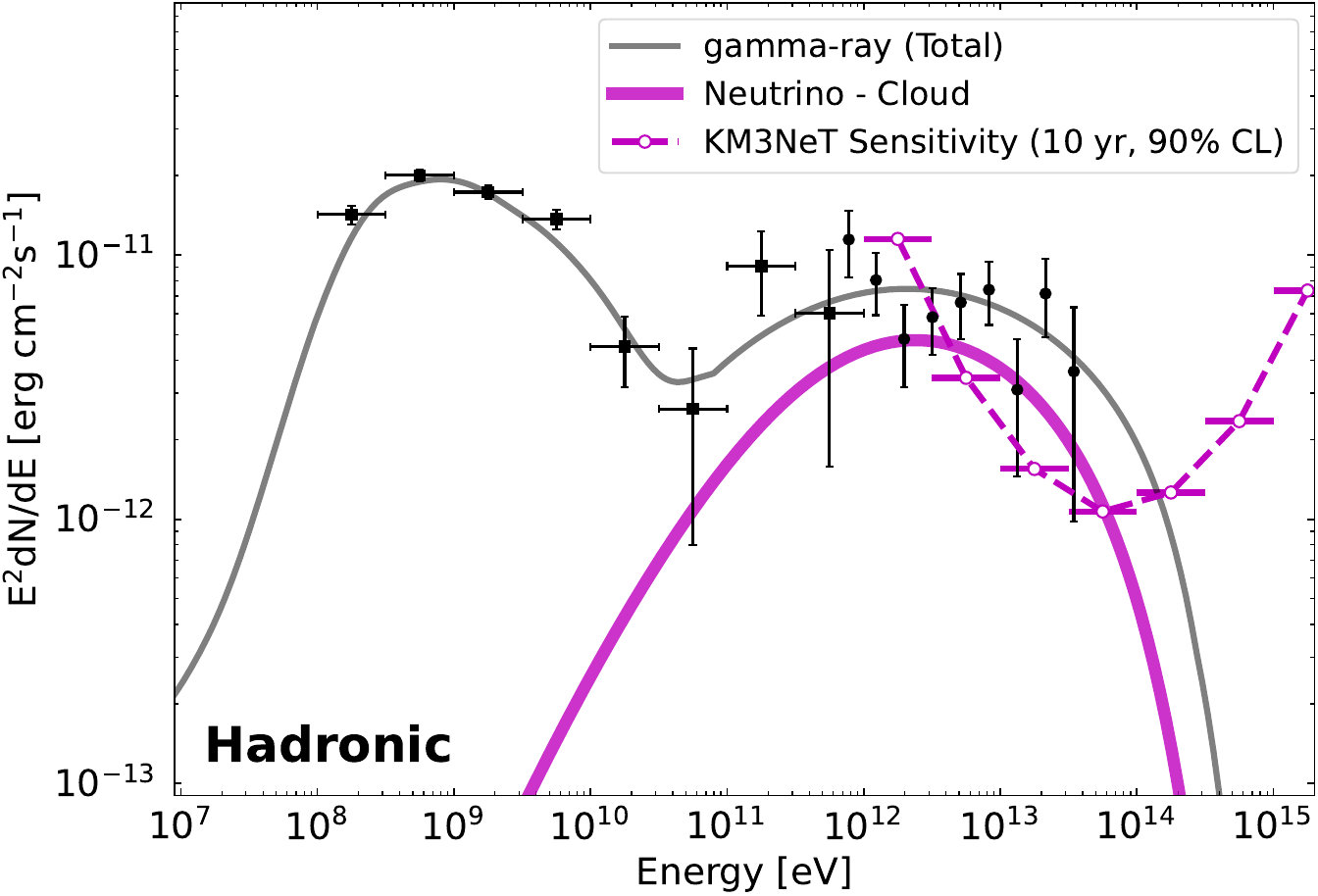}
        \end{minipage}
        \\
        \begin{minipage}{0.95\hsize}
            \hspace*{-0.5cm}\includegraphics[width=\hsize]{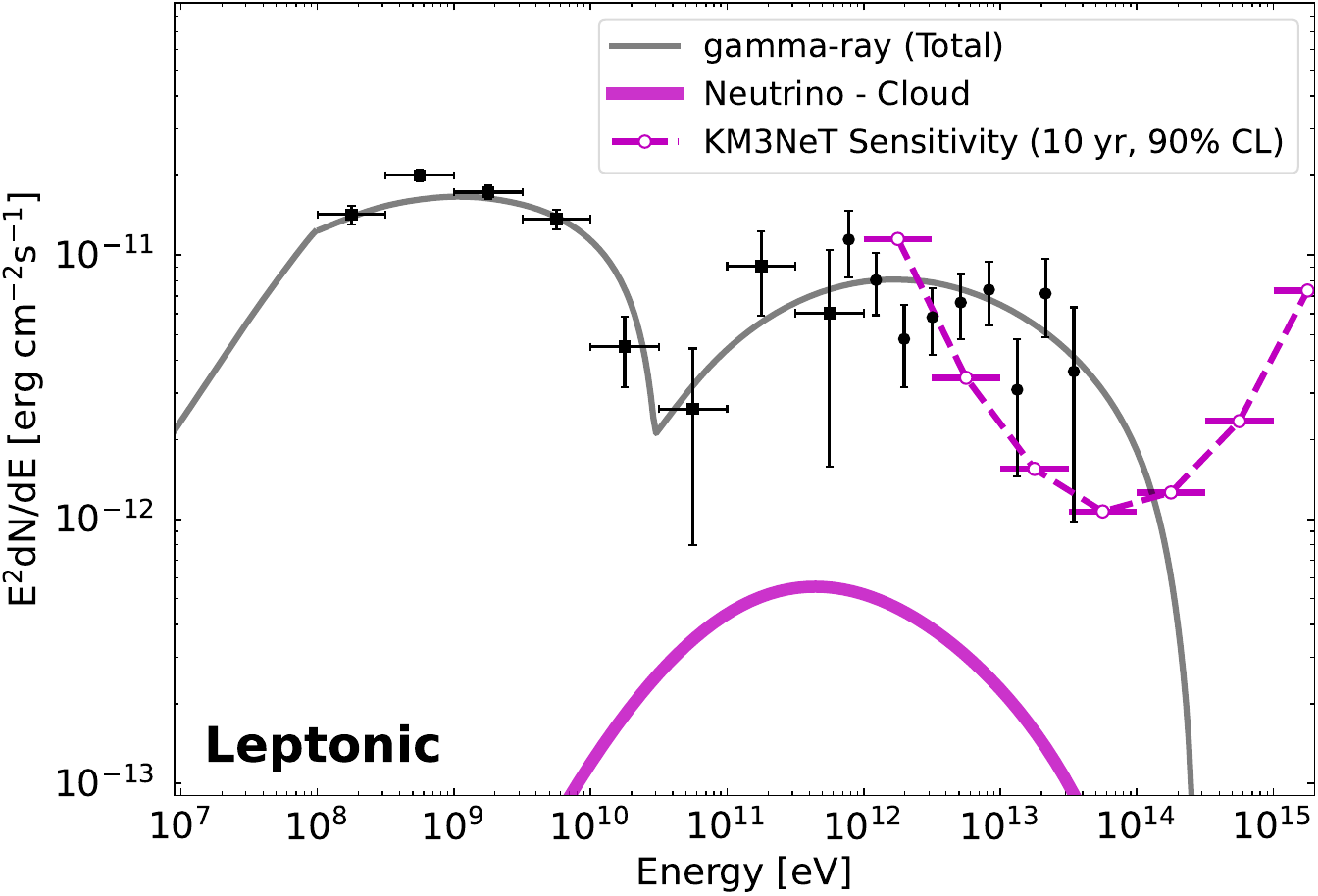}
        \end{minipage}
    \end{tabular}
    \caption{The predicted neutrino spectrum at the H.E.S.S. source for the hadronic-dominated (top) and leptonic-dominated case (bottom).
    The dashed line indicates the $90\%$ CL sensitivity of the KM3NeT experiment \cite{KM3NeT2024APh}.
    For comparison, the black points and black curves show the gamma-ray data and model spectra, respectively, which are the same as in Fig.~\ref{fig:modeling:best}.}
    \label{fig:modeling:neutrino}
\end{figure}

Future observations of high-energy neutrinos, e.g., KM3NeT \citep{KM3NeT2024APh} and TRIDENT \citep{TRIDENT2023NatAs}, can strengthen this claim.
We calculate neutrino spectra using the \verb|AAfrag| package \citep{Koldobskiy2021PhRvD}.
Fig.~\ref{fig:modeling:neutrino} shows the predicted neutrino spectrum at the cloud region for both hadronic and leptonic-dominated cases.
For the hadronic-dominated model, the neutrino flux at 1--100~TeV is compatible with the $90\%$ confidence level (CL) sensitivity of KM3NeT \citep{KM3NeT2024APh}\footnote{The detection sensitivity of the KM3NeT experiment depends on declination; here, we show the values at Dec = $-65.3^{\circ}$, which is the closest to the coordinates of HESS J1626$-$490 in the published sensitivity curves.}, while the flux in the leptonic-dominated case is much lower than the sensitivity.
Therefore, future observation in neutrinos will provide a critical test for the hadronic-dominated model.

\subsection{Constraints on maximum energy of CRs and diffusion coefficient}

\begin{figure}
    \begin{tabular}{c}
         \begin{minipage}{0.95\hsize}
            \centering
            \hspace*{-1cm}\includegraphics[width=\hsize]{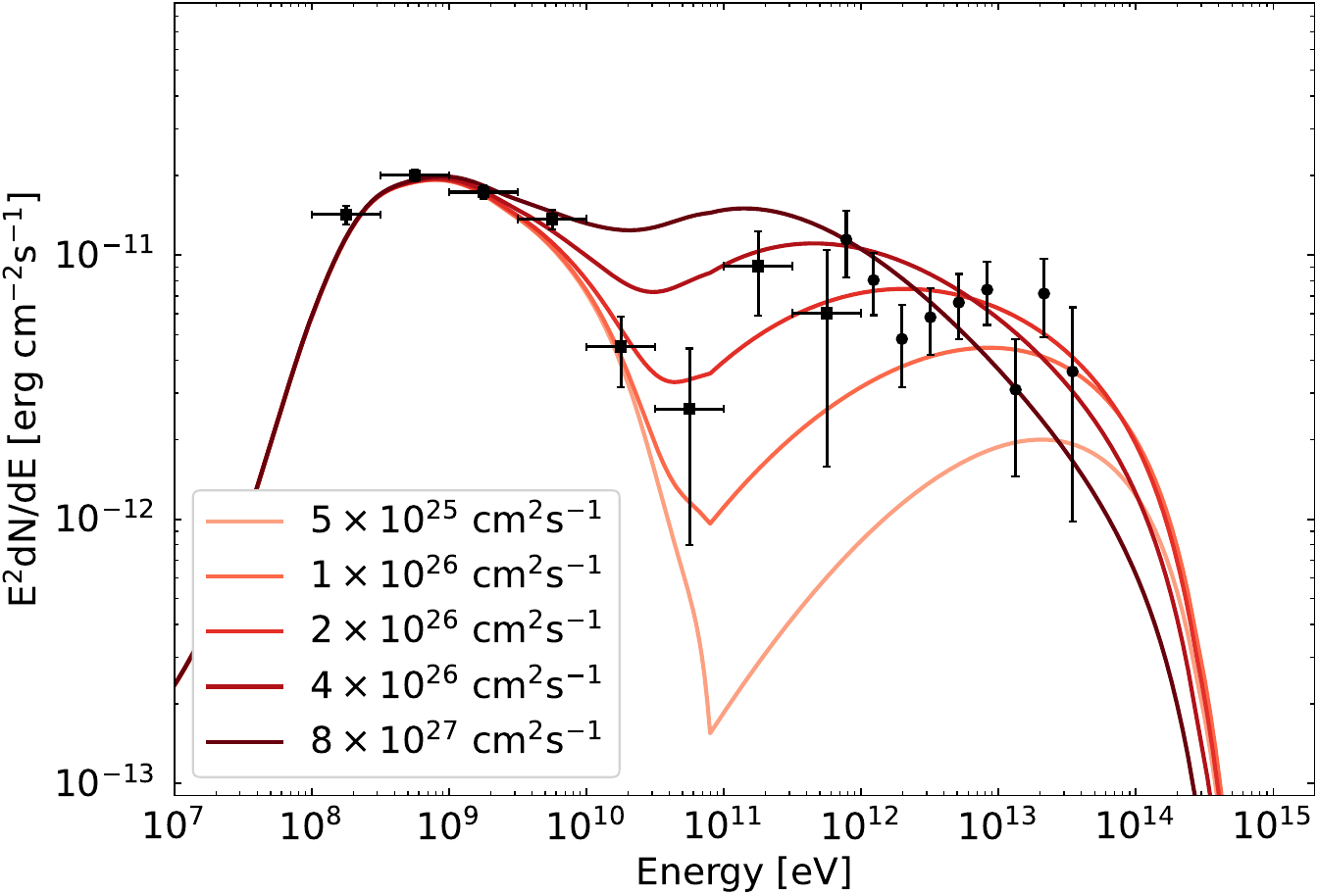}
        \end{minipage}
        \\
        \begin{minipage}{0.95\hsize}
            \centering
            \hspace*{-1cm}\includegraphics[width=\hsize]{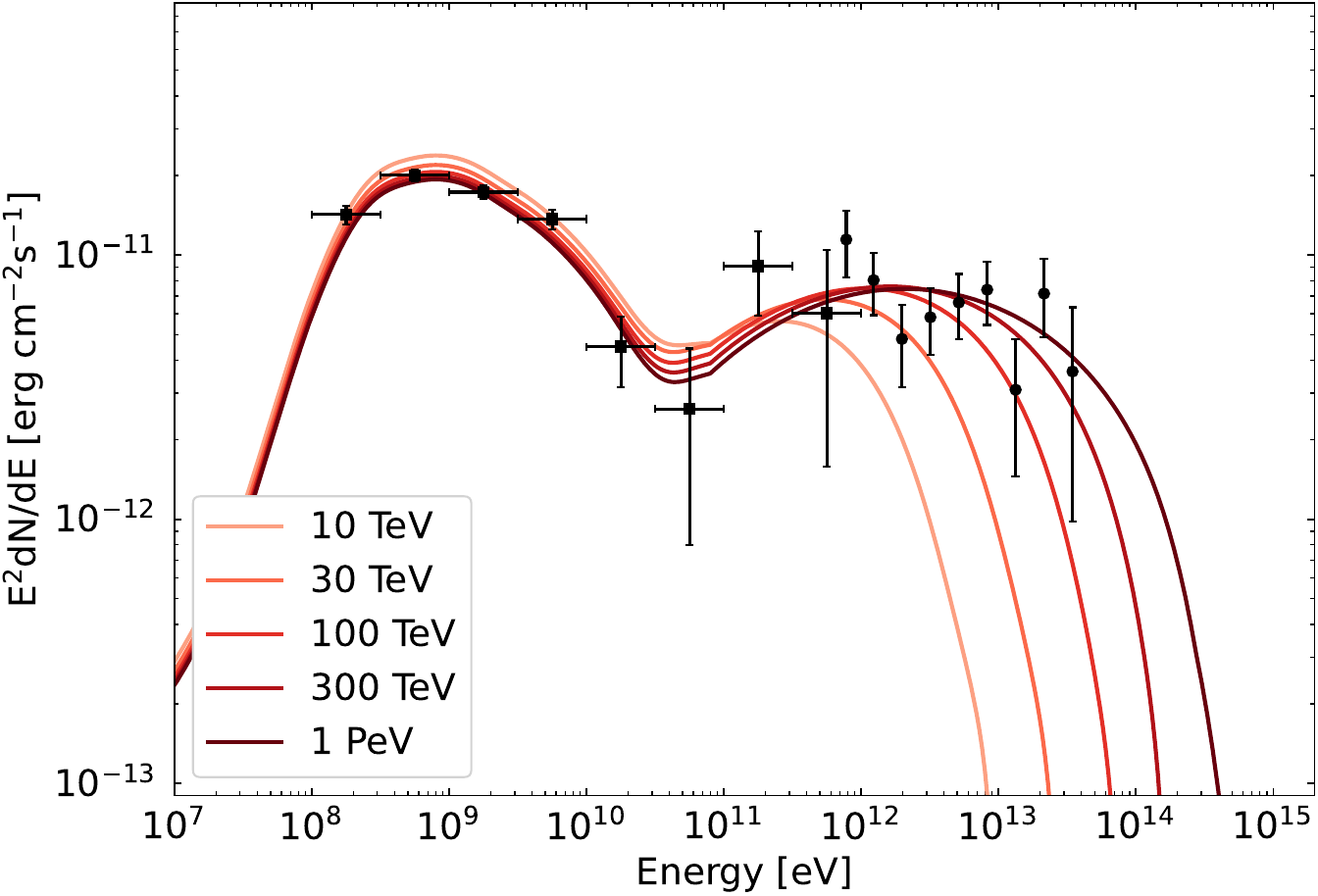}
        \end{minipage}
  \end{tabular}
\caption{The model parameters, $D_{0}$ (top) and $E_{\rm max}$ (bottom), dependency of the gamma-ray spectrum in the hadronic-dominated case. 
The data points in the gamma-ray bands are the same as in Fig.~\ref{fig:modeling:best}.
The values used for each parameter are written in the legend of the figure.}
\label{fig:modeling:modeldep}
\end{figure}

\begin{figure}
    \centering
    \includegraphics[width=0.95\hsize]{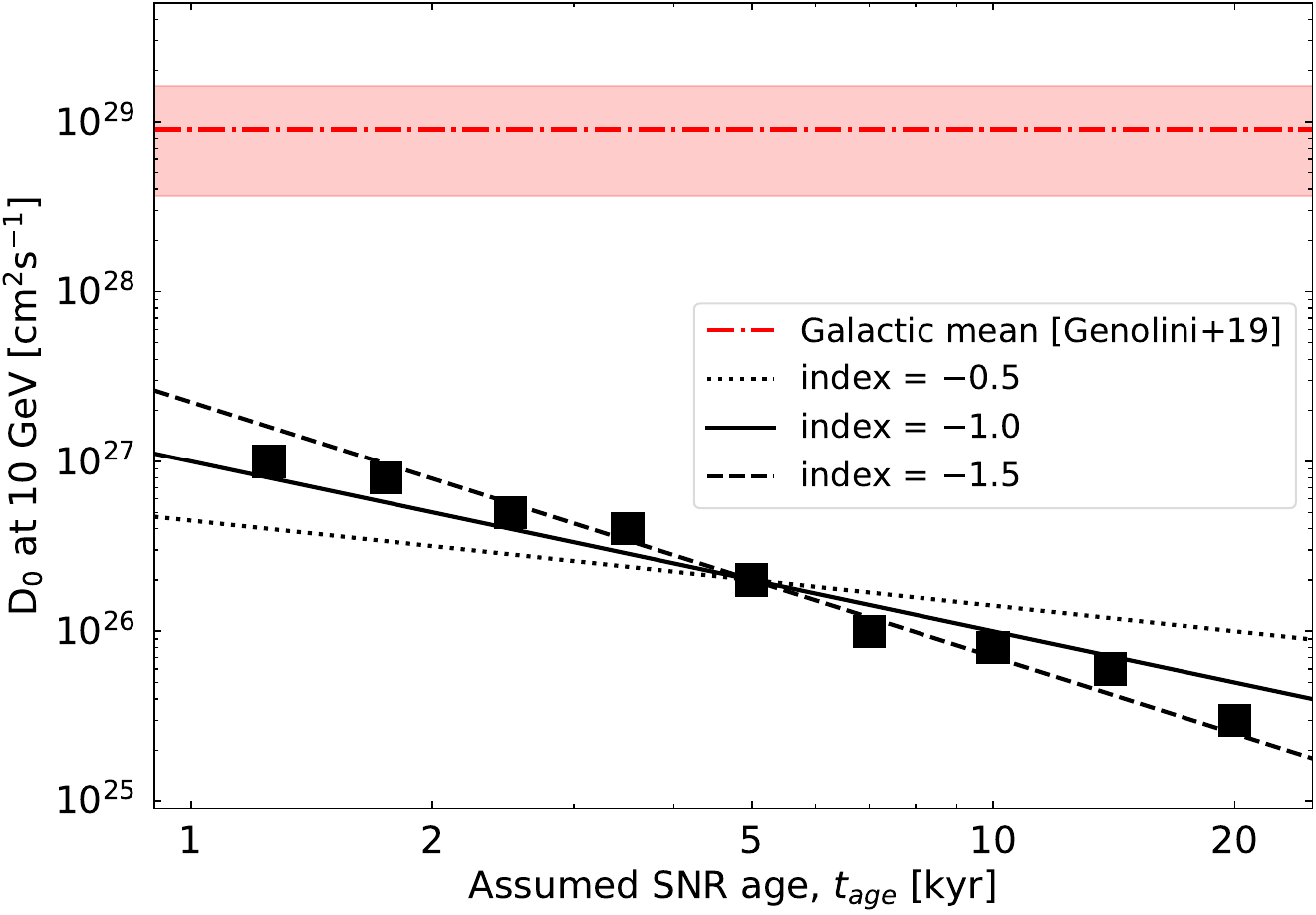}
\caption{The correlation between the diffusion coefficient ($D_{0}$) and assumed SNR age ($t_{\rm age}$).
    The data points show the best-fit parameters obtained with each SNR age.
    The dotted, solid and dashed lines show a power-law function of the assumed SNR age (normalized to $D_{0} = 2 \times 10^{26} ~\rm{cm^{2} s^{-1}}$ at $t_{\rm age} = 5~\rm{kyr}$) with an index of 0.5, 1.0, and 1.5, respectively.
    The red line and shaded area show the Galactic mean value derived from the CR spectrum observed on Earth and its uncertainty, the latter of which considers the minimum and maximum values derived for different Galactic halo sizes between 4 and 18~kpc in addition to statistical error in the fit \citep{Genolini2019PhRvD}.
}
\label{fig:modeling:d0_agedep}
\end{figure}

The diffusion coefficient ($D_{0}$) and maximum CR energy ($E_{\max}$) have a significant impact on the model calculations.
Fig.~\ref{fig:modeling:modeldep} shows the model variations with each parameter in the hadronic-dominated case.
The diffusion coefficient to reproduce the observed spectrum is found to be $D_{0} \sim 2 \times 10^{26}~\rm{cm^{2}s^{-1}}$, which is about two orders smaller than the Galactic mean (${\sim} 10^{29}~\rm{cm^{2}s^{-1}}$; e.g., \citet{Cummings2016ApJ}, \citet{Genolini2019PhRvD}).
The estimate of $D_{0}$ depends on an assumed SNR age ($t_{\rm age}$).
Fig.~\ref{fig:modeling:d0_agedep} shows the correlation between $D_{0}$ and $t_{\rm age}$, indicating that $D_{0}$ is inversely proportional to $t_{\rm age}$.
According to \cite{Suzuki2021}, an estimate from a dynamical age, as in this study, agrees with an actual age within a factor of 4.
If the actual age of the SNR is four times smaller, the diffusion coefficient will be estimated as $D_{0} \sim 1 \times 10^{27}~\rm{cm^{2}s^{-1}} (t_{\rm age}/1.2~\rm{kyr})$, which is still lower than the Galactic mean.
Such a small diffusion coefficient is explained in conjunction with CR self-confinement caused by the generation of turbulent plasma waves \citep[e.g.,][]{Wentzel1974, Fujita2011MNRAS, DAngelo2017} and also seen in other SNRs \citep[e.g.,][]{Fujita2009, Ohira2010_MNRAS, Li2012MNRAS}.
Interestingly, the diffusion coefficient varies with SNR.
For example, the diffusion coefficient around SNR HB9 is close to the Galactic mean, implying no slow diffusion in its vicinity \citep{Oka2022, Bao2024ApJ}.
A simple explanation for the diversity of the diffusion coefficient may be differences in CR density; the CR energy density in the H.E.S.S. source region is ${\sim} 1.0~\rm{eV\,cm^{-3}}$, which is similar to that of HB9 (0.3--1.1~$\rm{eV\,cm^{-3}}$).
Differences in the magnetic field and/or the plasma state can also drive slow diffusion \citep[e.g.,][]{Fujita2011MNRAS, Malkov2013ApJ, DAngelo2017}, but these discussions are beyond the scope of this paper.
The evidence for slow diffusion around the SNR, as with the situation around PWNe \citep[e.g.,][]{HAWC2017Sci}, will have implications for studies of modern CR propagation \citep[e.g.,][]{Cummings2016ApJ, Johannesson2016ApJ} and Galactic diffuse emission \citep{Fermi2012ApJ, Orlando2018MNRAS,Amenomori2021PhRvL, Cao2023PhRvL}.

Fig.~\ref{fig:modeling:modeldep} (b) suggests that the model requires $E_{\max}$ of ${\gtrsim}100$~TeV to reproduce the H.E.S.S. spectrum above 10~TeV.
Future observations with the new generation of gamma-ray observatories, e.g., CTA-South \citep{CTA2013APh}, ALPACA \citep{ALPACA2021ExA}, and SWGO \citep{SWGO2019}, will evaluate an energy cutoff at $>10$~TeV, enabling us to precisely determine $E_{\max}$ and elucidate whether this SNR is a PeV CR accelerator.

\subsection{Alternative scenarios}

One can consider the contribution of a PWN as an alternative scenario for the gamma-ray origin. 
We estimate the amount of gamma-ray emission from the nebula of PSR J1627$-$4845.
With the spin-down luminosity ($L_{\rm sd}$) of the pulsar and the distance from the Earth ($D$), the gamma-ray flux from a PWN is approximated as $F \sim L_{\rm sd} \eta_{\gamma} / 4 \pi D^{2}$, where $\eta_{\gamma}$ is the conversion efficiency from the spin-down energy to the radiation \citep{Fermi2011ApJ_PWN, HESS2018A&A_PWN}.
Instituting $L_{\rm sd} = 6.3 \times 10^{32}~{\rm erg~s^{-1}}$ \citep{Kaspi1996} and $d = 3.3 \pm 0.6~{\rm kpc}$, we obtain $F \sim 4.8_{-1.4}^{+2.4} \times 10^{-14} \times (\eta_{\gamma}/0.1)~{\rm erg~cm^{-2}s^{-1}}$, indicating that even an optimistic estimate with $\eta_{\gamma}=0.1$ gives flux far below the observed one ($F_{\gamma} > 10^{-12}~{\rm erg~cm^{-2}s^{-1}}$).

Additionally, another pulsar PSR J1625$-$4904 is located $0.29^{\circ}$ away from the H.E.S.S. source.
This pulsar has a higher spin-down luminosity of $6.7 \times 10^{33}~{\rm erg~s^{-1}}$ than that of PSR J1627$-$4845, but the distance from the Earth is 7.9~kpc \citep{Manchester2001MNRAS}.
The predicted gamma-ray flux is $F \sim 9.1 \times 10^{-14}~{\rm erg~cm^{-2}s^{-1}}$ and thus very much smaller than the observed flux.

According to \citet{HESS2018A&A_PWN}, PWNe with TeV detections always have a high spin-down luminosity of $\gtrsim 10^{35}~\rm{erg\,s^{-1}}$, which is more than an order of magnitude larger than those of the above-mentioned pulsars.
We thus conclude that a PWN of a known pulsar is unlikely to be the gamma-ray origin.

Note that we do not model the spectrum of point source 4FGL J1629.3$-$4822c in this paper, while we briefly discuss its origin here.
This is one of the unidentified GeV sources as listed in the catalog \citep{FermiLAT2019}.
In the vicinity of the point source (within a radius of $0.1^{\circ}$), there are two AGN candidates \citep{Edelson2012ApJ}, seven dark clouds \citep{Peretto2016A&A}, and the radio compact source G335.69+00.20 \citep{Planck2014A&A}.
Although the counterpart is potentially among them, no clear association has been seen in the recent studies by cross-matching catalogs \citep{Bhat2022A&A, Ajello2022ApJS}.
Another possibility is due to escaped CRs from SNR G335.2+0.1, as in the case of the extended source Gaussian-W.
Although we assumed a simple morphology for GeV gamma-ray emission, symmetric Gaussian, it may be more complicated.
The recently released catalogue from the Fermi-LAT Collaboration \citep[4FGL-DR4;][]{Ballet2023arXiv} includes a new unknown point source, 4FGL J1626$-$4858c, detected with $4.8\sigma$ confidence level in this region.
In this work, we focused only on the emission corresponding to the SNR and H.E.S.S. sources but would need to consider spatial models that better describe all the emissions in this vicinity in the future.

\section{Summary} \label{section:Summary}

To resolve the origin of the unidentified TeV gamma-ray source HESS J1626$-$490, we scrutinized the properties of the supernova remnant G335.2+0.1 in the vicinity of the H.E.S.S. source and searched for non-thermal counterparts by analyzing multi-wavelength data in the radio, X-ray, and GeV gamma-ray bands.
The $^{12}$CO ($J{=}1{-}0$) and H\,{\sc i} line data has shown two cavity-like structures coinciding with the SNR at radial velocity ranges of $V_{\rm LSR} = -22~\rm{km\,s^{-1}}$ and $-45~\rm{km\,s^{-1}}$.
We examined the absorption line features and find that one ($V_{\rm LSR} = -22~\rm{km\,s^{-1}}$) is due to absorption by radio continuum radiation from the SNR, while the other ($-45~\rm{km\,s^{-1}}$) is indeed associated with the SNR.
The distance to the SNR has been then estimated to be $3.3 \pm 0.6$~kpc.
We also analyzed X-ray and GeV gamma-ray data from the vicinity of this region, observed with the XMM-Newton and \textit{Fermi}-LAT.
Although no X-ray counterpart was found, we have detected an extended GeV gamma-ray emission with a Gaussian $\sigma$ of $0.31\pm0.01^{\circ}$, overlapping with both the SNR and H.E.S.S. source regions.
The location of the GeV emission changes with energy and is in good agreement with the SNR in the lower energy range below 2~GeV, while in the higher energy range above 11~GeV, it agrees with the H.E.S.S. source.
We also found the energy spectrum has a spectral hardening at $\sim 50$~GeV, indicating the existence of two components in the gamma-ray emission.
We modeled the observed spectrum by combining emissions from the SNR and the cloud regions, assuming that CRs were accelerated in the SNR and illuminated the cloud emitting gamma rays.
We then found both the leptonic and hadronic emissions from the cloud can explain the TeV gamma-ray spectrum.
However, the leptonic model requires an unrealistic assumption about the time evolution of the magnetic field at the SNR (specifically, the time when CR electron starts to escape from the SNR), suggesting that the hadronic model is preferred.
Our hadronic model suggests that the diffusion coefficient around the SNR is two orders of magnitude lower than the Galactic mean. 
This small diffusion coefficient around the SNR indicates a limitation in the modern CR propagation model with isotropic diffusion, highlighting the importance of further investigating the diversity of diffusion coefficients.
We have also shown that this SNR has accelerated protons up to ${\gtrsim}100$~TeV in the past.
Future southern observations of sub-PeV gamma rays and neutrinos can provide a critical test of whether the SNR G335.2+0.1 was a PeVatron.

\begin{acknowledgments}
We appreciate the anonymous referee for valuable comments.
The \textit{Fermi} LAT Collaboration acknowledges generous ongoing support from a number of agencies and institutes that have supported both the development and the operation of the LAT as well as scientific data analysis.
These include the National Aeronautics and Space Administration and the
Department of Energy in the United States, the Commissariat \`a l'Energie Atomique and the Centre National de la Recherche Scientifique / Institut National de Physique Nucl\'eaire et de Physique des Particules in France, the Agenzia Spaziale Italiana
and the Istituto Nazionale di Fisica Nucleare in Italy, the Ministry of Education, Culture, Sports, Science and Technology (MEXT), High Energy Accelerator Research Organization (KEK) and Japan Aerospace Exploration Agency (JAXA) in Japan, and the K.~A.~Wallenberg Foundation, the Swedish Research Council and the Swedish National Space Board in Sweden.
This work was supported by a Grant-in-Aid for JSPS Fellows Grant No. 23KJ2094 (TO) and JSPS KAKENHI grant No. 24H00246 (HS).
\end{acknowledgments}

\vspace{5mm}
\software{astropy \citep{astropy2013, astropy2018, astropy2022}, APLpy \citep{aplpy2012, aplpy2019}, matplotlib \citep{Hunter2007}, numpy \citep{Harris2020}, gammapy \citep{gammapy2023}}

\appendix

\section{\textit{Fermi}-LAT spectral analysis with the double-Gaussian model} \label{section:AppendixA}

\begin{figure}
    \begin{tabular}{cc}
        \begin{minipage}{0.47\hsize}
            \hspace*{-0.5cm}\includegraphics[width=\hsize]{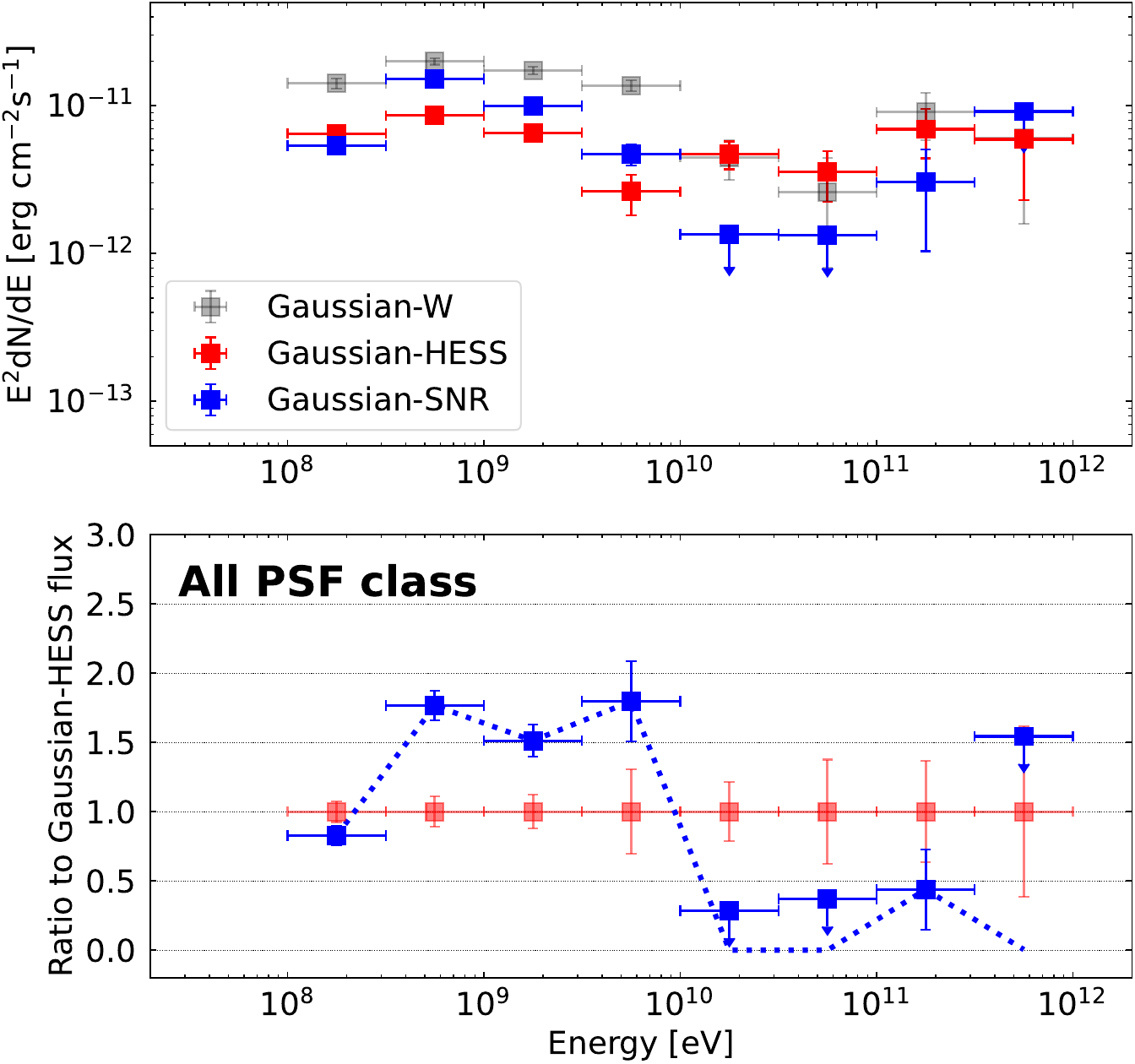}
        \end{minipage}
        \begin{minipage}{0.47\hsize}
            \hspace*{-0.5cm}\includegraphics[width=\hsize]{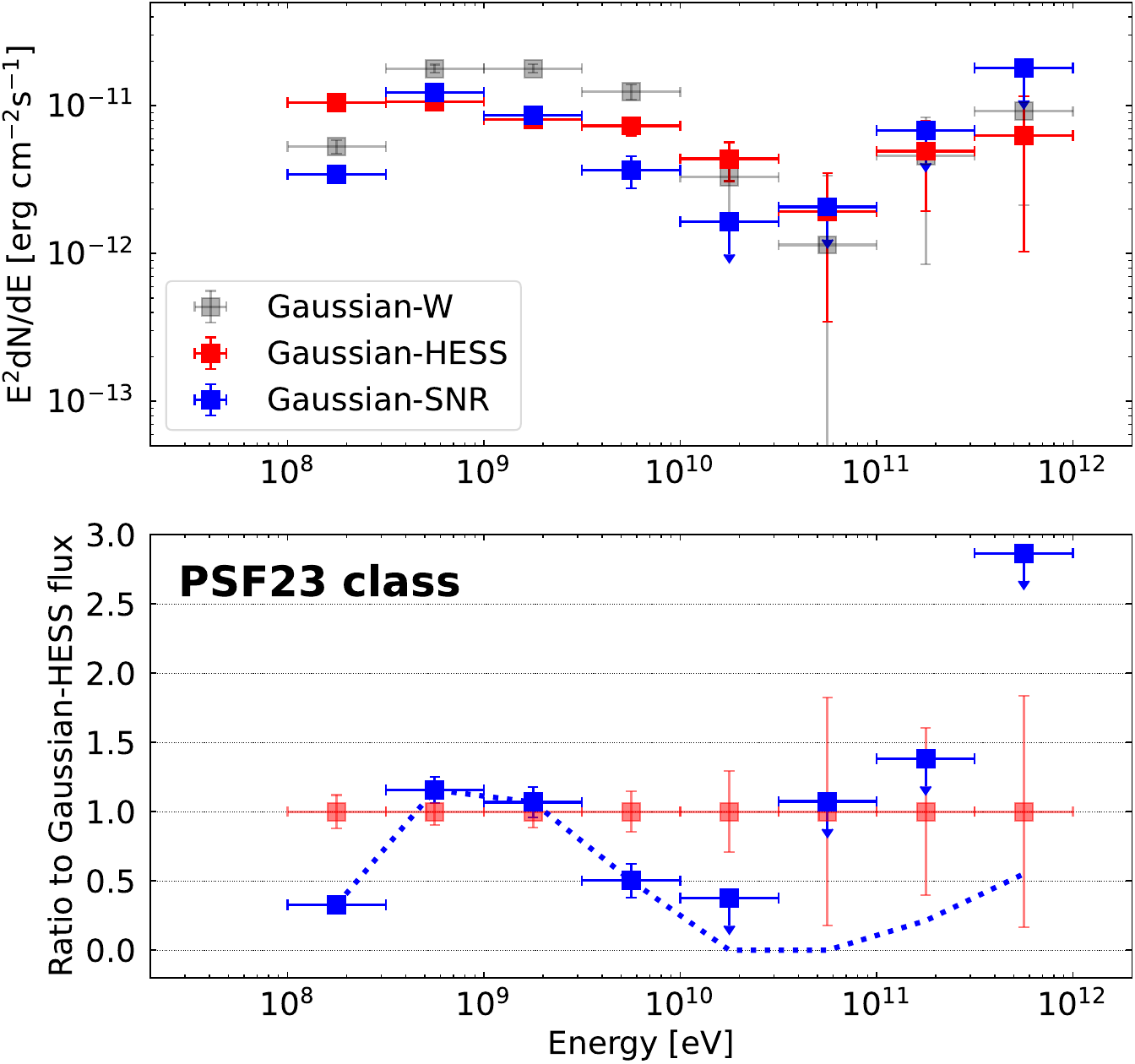}
        \end{minipage}
    \end{tabular}
    \caption{Gamma-ray spectra obtained with the 2-Gaussian model.
    Left panels show the results with all PSF class data as in the main text, while right panels show the results when only PSF2 and PSF3 classes are chosen.
    The blue and the red markers show the spectrum of the Gaussian sources at SNR G335.2+0.1 and at HESS J1626$-$4920, respectively.
    In the top panels, the gray points are the results for Gaussian-W.
    The bottom panels show the flux ratio of the source at SNR to that at HESS J1626$-$4920.
    The meaning of the markers is the same as in the top panel.
    The dotted blue line connects the flux obtained at each energy bin to guide the eye.
    }
    \label{fig:Fermi_Spectrum:DoubleGauss}
\end{figure}

The likelihood ratio test (Sect. \ref{sec:ana:fermi}) indicates that the 2-Gaussian model, which consists of two extended sources with the same spatial parameters as SNR G335.2+0.1 and HESS J1626$-$4920, is disfavored compared to the 1-Gaussian model.
However, as shown in Fig. \ref{fig:GeVmap} and \ref{fig:Fermi_GaussFit}, the energy dependence of the morphology suggests the presence of multiple radiation components, and in particular, the radiation from the SNR and H.E.S.S. sources appears to be dominant.
Therefore, it is meaningful to obtain the spectra of these regions separately.
Fig. \ref{fig:Fermi_Spectrum:DoubleGauss} shows the energy spectra of the two sources with the 2-Gaussian model.
In the high-energy band above 10~GeV, there is a visible difference between the two sources, and the SNR region has significantly lower flux than the H.E.S.S. source region.
In contrast, the spectra below 10~GeV have similar shapes because the deterioration of angular resolution prohibits resolving the two objects.
Here, we perform the same analysis but with data of PSF2 and PSF3 classes, which have relatively good angular resolution.
The results are shown in Fig. \ref{fig:Fermi_Spectrum:DoubleGauss} (b).
The spectral shapes are still similar to the ones obtained with all PSF classes combined.

\begin{figure}
    \begin{tabular}{cc}
        \begin{minipage}{0.47\hsize}
            \hspace*{-0.5cm}\includegraphics[width=\hsize]{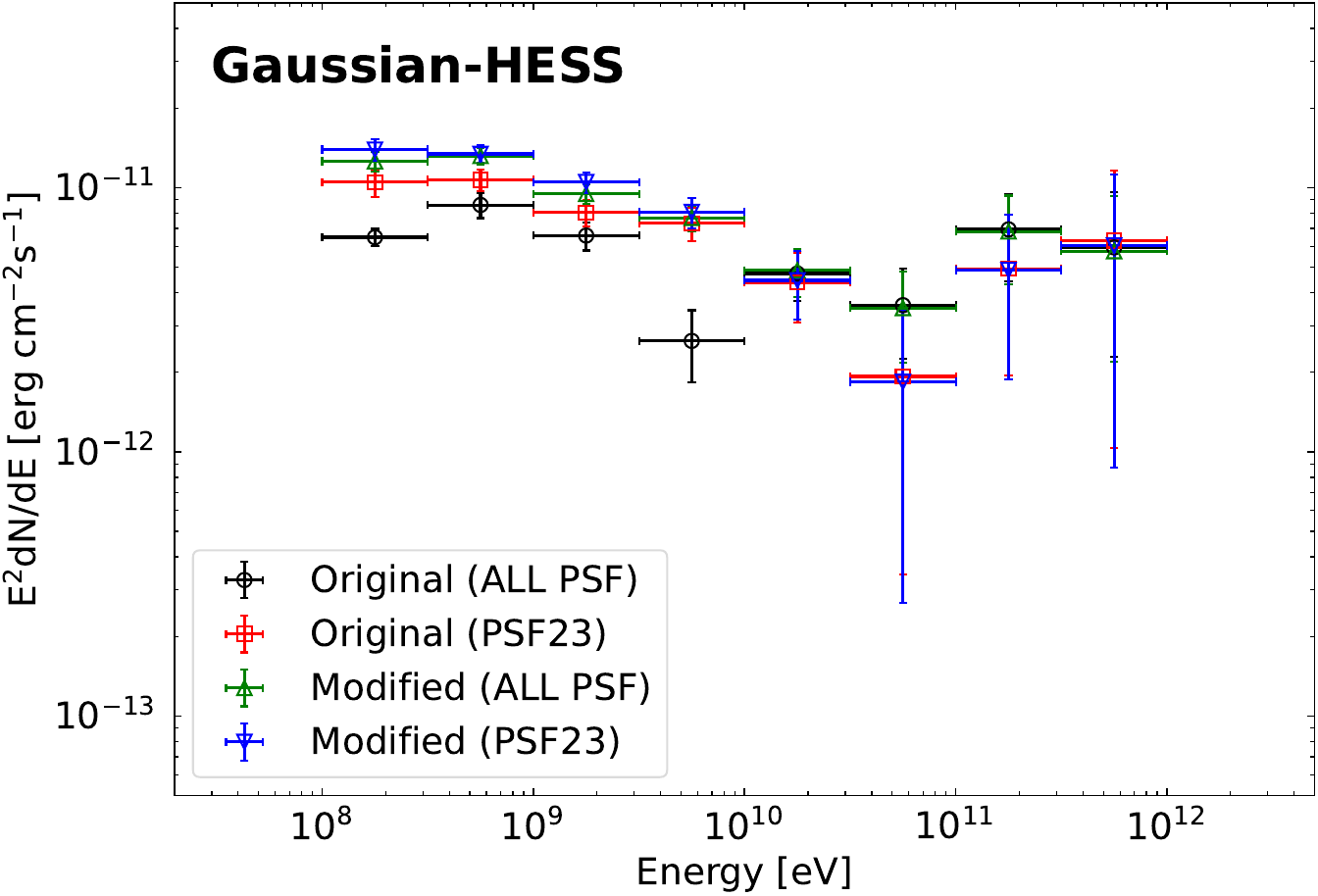}
        \end{minipage}
        \begin{minipage}{0.47\hsize}
            \hspace*{-0.5cm}\includegraphics[width=\hsize]{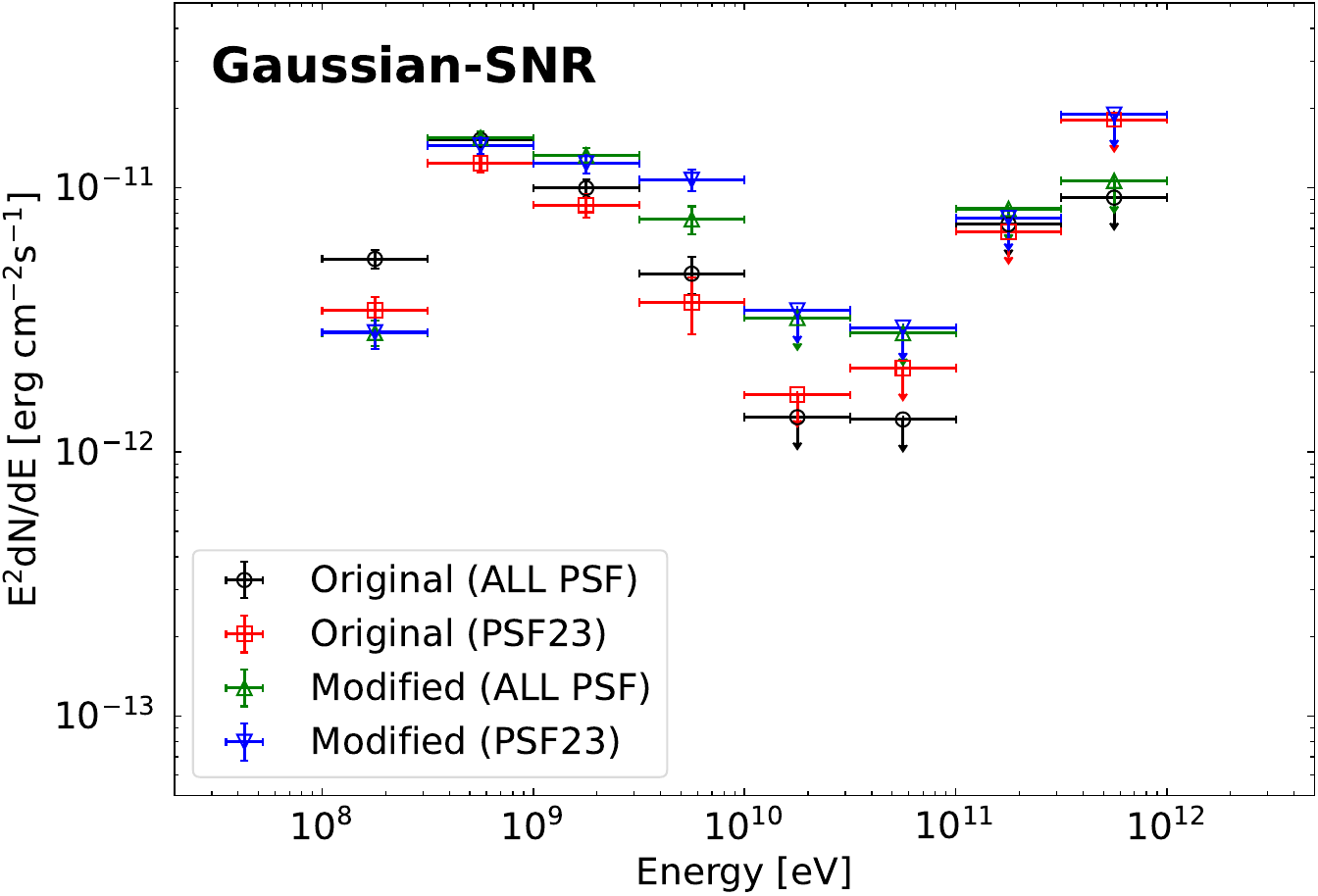}
        \end{minipage}
    \end{tabular}
    \caption{Comparison of energy spectra between the modified 2-Gaussian model and the original model.
    The left and right panels show the Gaussian source at HESS J1626$-$490 and SNR G335.2+0.1, respectively.
    The black circles (red squares) show the result when the original model is applied to all PSF (PSF2 and 3) class data, while the green triangles (blue inverted triangles) show the results when the modified model is applied to all PSF (PSF2 and 3) class data.
    }
    \label{fig:Fermi_Spectrum:DoubleGauss_free}
\end{figure}

Thus far, we have adopted values for the parameters of the 2-Gaussian model based on observational data at other wavelengths, but these are not optimized for the GeV morphology.
Here, we optimize the spatial parameters of the 2-Gaussian model as follows.
First, we scan the Gaussian parameters of the SNR source to maximize the likelihood value, while fixing the H.E.S.S. source.
We then scan the parameters for the H.E.S.S. source fixing the SNR source obtained above.
This procedure is repeated until the resulting parameters match the previous values within $1\sigma$ errors.
The obtained likelihood value, the difference from the null hypothesis, is $\Delta TS = 888.3$, which is comparable to the 1-Gaussian+1-PS model.
The best-fit Gaussian parameters are (RA, DEC, $\sigma$) = ($246.73 \pm 0.04^{\circ}$, $-49.15 \pm 0.02^{\circ}$, $0.20 \pm 0.01^{\circ}$) for the H.E.S.S. source and ($247.40 \pm 0.04^{\circ}$, $-48.60 \pm 0.02^{\circ}$, $0.22 \pm 0.02^{\circ}$) for the SNR source.
While the parameters of the H.E.S.S. source are almost the same as the original values, the SNR source changes significantly.
Fig.~\ref{fig:Fermi_Spectrum:DoubleGauss_free} shows the energy spectra of the two Gaussian sources.
The trends remain the same as in the original case, suggesting that the two emissions are not resolved at the low energy band.

\section{CR-escape model} \label{sec:appendix:cr_escape_model}

Generally, the maximum particle energy in an SNR depends on the timescale of escape from the acceleration region or energy losses due to radiative cooling \citep[e.g.,][]{Ptuskin2005A&A}.
In the case where energy losses are negligible (e.g., protons), we can assume the phenomenological power-law dependence on age $t$ for the cutoff energy $E_{\rm esc}$ of the CR spectrum at the SNR shock \citep[e.g.,][]{Gabici2009, Ohira2010A&A}:
\begin{equation} \label{eq:Ecut}
E_{\rm esc}(t) = E_{\max}\left(\frac{t}{t_{\rm Sedov}}\right)^{-\alpha},
\end{equation}
where $t_{\rm Sedov}$ is the time at which the SNR has entered the Sedov phase, and $E_{\max}$ is the maximum particle energy at $t_{\rm Sedov}$.
Assuming that the initial velocity of the blast wave is $u_{\rm sh,0} = 10^{9}~\rm{cm\,s^{-1}}$ and $E_{\rm SN}$ and $n$ are $0.3 \times 10^{51}$~erg and $0.3~\rm{cm^{-3}}$, respectively, as in Eq.~\ref{eq:t_age}, we obtain $t_{\rm Sedov} = 210$~yr.
The power-law index ($\alpha$) is determined once we have $E_{\rm max}$ and the current maximum energy ($E_{\rm now}$), which are deduced from fitting to the observed spectra.
The timescale for a particle with an energy of $E$ to escape into interstellar space from the supernova explosion is given by 
\begin{equation}\label{eq:t_esc}
t_{\rm esc}(E)=t_{\rm Sedov}\left(\frac{E}{E_{\max}}\right)^{-1/\alpha}.
\end{equation}
For electrons, we need to take into account the radiative cooling and parameterize the starting time of escape into the interstellar space as follows: 
\begin{equation} \label{eq:xi_e_inj}
t_{\rm e} = \xi_{\rm e, inj} t_{\rm Sedov},
\end{equation}
where $\xi_{\rm e, inj}$ is a scaling factor and in the limit of $\xi_{\rm e} \rightarrow 1$, the injection is identical to that of protons.
After evaluating $t_{\rm e}$, we use Eqs. \ref{eq:Ecut} and  \ref{eq:t_esc} for electrons, as we expect the time evolution of the maximum energy is the same as that of protons.

To obtain the distribution function $f_{i, \rm out}$ per unit energy per unit volume of the CR of nuclide $i$ escaping into interstellar space, we solve the transport equation with radiative cooling,
\begin{equation} \label{eq:transport}
   \frac{\partial f_{i,\rm out}}{\partial t}(t, r, E)
    - D_{\mathrm{ISM}}(E) \Delta f_{i,\rm out}(t, r, E)
    +\frac{\partial}{\partial E}\left(P(E)f_{i,\rm out}(t, r, E)\right)
    = q_{i, s}(t, r, E),
\end{equation}
where $r$ is the distance from the SNR center, $P(E)$ is an energy loss rate (which is ignored for protons), $D_{\mathrm{ISM}}$ is the diffusion coefficient in the ISM, and $q_{i, s}$ is the injection rate of the CR from the SNR shock into interstellar space per unit energy, unit volume, and unit time.
As for the cooling process, we consider the synchrotron radiation, which is the most dominant effect for electrons with energies above $\mathcal{O}({\rm GeV})$ \citep[e.g.,][]{Berezinskii1990book}.
We adopt a simple power-law form for the diffusion coefficient:
\begin{equation}\label{eq:DISM}
D_{\mathrm{ISM}}(E) = D_0 \left(\frac{E}{10~{\rm GeV}}\right)^{\delta},
\end{equation}
where $D_0$ is the diffusion coefficient of CRs at $E = 10~{\rm GeV}$.
We allow $D_0$ to vary freely because this parameter may be modified by the effect of CR-self confinement caused by the generation of turbulent plasma waves \citep[e.g.,][]{Wentzel1974}, while we fix $\delta = 1/3$, as expected in the Kolmogorov turbulence theory \citep{Kolmogorov1991}.
If a larger $\delta$ is adopted, the estimate of $D_0$ will be smaller \citep[e.g.,][]{Oka2022}. For example, in the hadronic-dominated model, $D_0$ is estimated to ${\sim}1 \times 10^{26}~{\rm cm^{2} s^{-1}}$ for $\delta = 0.5$ and ${\sim}2 \times 10^{25}~{\rm cm^{2} s^{-1}}$ for $\delta = 0.7$. This fact strengthens our claim of a slow diffusion around SNR G335.2+0.1.

According to \cite{Ohira2010_MNRAS}, CRs with a monochromatic energy of $E$ are injected from the SNR shell with a radius of $R_{\rm esc} = u_{\rm sh,0} t_{\rm Sedov} (t_{\rm esc}(E)/t_{\rm Sedov})^{2/5}$ at $t = t_{\rm esc}(E)$.
The respective injection rates for protons $q_{p, s}$ and electrons $q_{e, \rm s}$ are then given by
\begin{eqnarray} \label{eq:qs}
    q_{p, s} &=& \frac{N_{\mathrm{esc}}(E)}{4\pi r^2} \delta(r-R_{\rm esc}(E)) \delta\left(t-t_{\rm esc}(E)\right), \\
    \label{eq:qs_e}
    q_{e,s} &=& K_{\rm ep} q_{p,s},
\end{eqnarray}
where $r$ is the distance from the center of the SNR, $N_{\rm esc}$ is the spectrum of all CRs that have escaped from the SNR until $t_{\rm age}$, and $K_{\rm ep}$ is the electron-to-proton flux ratio.
Following \cite{Oka2022}, we assume the total energy of the escaped CRs is proportional to the supernova explosion energy, namely
\begin{equation}\label{eq:Nesc_normalize}
    \int_{E_{\rm now}}^{E_{\max}} E N_{\rm esc}(E) dE =\eta E_{\rm SN},
\end{equation}
where $\eta$ is the acceleration efficiency coefficient.
We also assume that $N_{\rm esc}$ is a power-law function of $E$ with an index $p_{\rm esc}$. 
\cite{Ohira2010A&A} suggested that $p_{\rm esc}$ is steeper than the index of CRs confined in the SNR shell ($p_{\rm SNR} \sim 2$); The propagation models \citep{Obermeier2012ApJ, Yuan2017PhRvD} expect the spectral index of particles escaped from Galactic CR origins is $p_{\rm esc} = 2.2$--$2.4$.
We adopt $p_{\rm esc} = 2.3$ as a fiducial parameter in model fitting.
The integral of Eq.~\ref{eq:Nesc_normalize} when $p_{\rm esc} \neq 2$ yields
\begin{equation} \label{eq:Nesc}
 N_{\rm esc}(E) = \frac{\eta \left(2-p_{\rm esc}\right) E_{\rm SN}}{E_{\rm now}^2} \left[\left(\frac{E_{\rm max}}{E_{\rm now}}\right)^{2-p_{\rm esc}}-1\right]^{-1}\left(\frac{E}{E_{\rm now}}\right)^{-p_{\rm esc}}.
\end{equation}

We obtain the energy spectrum of protons by combining the transport equation (Eq.~\ref{eq:transport}) and Eqs.~\ref{eq:qs} and \ref{eq:Nesc}:
\begin{equation} \label{eq:f_out}
f_{\rm{p,out}}(t, r, E) = \frac{N_{\rm {esc }}(E)}{4 \pi^{3 / 2} r R_{\rm {esc}} R_{\mathrm{d,p}}} \times \left[\exp \left(-\frac{\left(r-R_{\rm{esc}}\right)^{2}}{R_{\rm d,p}^{2}}\right)-\exp \left(-\frac{\left(r+R_{\rm{esc }}\right)^{2}}{R_{\rm d,p}^{2}}\right)\right],
\end{equation}
$R_{\rm d,p}$ is the diffusion length of protons, defined as
\begin{equation} \label{eq:RDiff}
R_{\rm d, p} \equiv \sqrt{4D_{\mathrm{ISM}}(E)\left(t-t_{\rm esc}(E)\right)}.
\end{equation}
The energy spectrum of electrons is obtained with a similar way as in protons but taking into account the synchrotron cooling.
According to \cite{Oka2022}, the spectrum of electrons is given by
\begin{equation}\label{eq:f_oute}
f_{\rm {e,out}}(t, r, E) = \frac{K_{\rm ep}N_{\mathrm{esc}}\left(E_{c}\right)}{4 \pi^{3 / 2} r R_{\mathrm{c}} R_{\rm d, e}} \frac{E_{c}^{2}}{E^{2}} \frac{1}{1-Q_{\rm syn}t_{c}E_c / \alpha} \times \left[\exp \left(-\frac{\left(r-R_{\mathrm{c}}\right)^{2}}{R_{\rm d, e}^{2}}\right)-\exp \left(-\frac{\left(r+R_{\mathrm{c}}\right)^{2}}{R_{\rm d, e}^{2}}\right)\right].
\end{equation}
The diffusion length of electrons is written as
\begin{equation} \label{eq:RDiff_e}
R_{\rm d, e} \equiv \sqrt{\frac{4D_{\rm ISM}(E)}{\left(1-\delta\right) Q_{\rm syn}E}
\left[1-\left(\frac{E}{E_c}\right)^{1-\delta}\right]},
\end{equation}
where $Q_{\rm syn}=\sigma_T B_{\rm ISM}^2/\left(6\pi m_e^2 c^3\right)$ is the efficiency of synchrotron radiation and is proportional to the square of the magnetic field in the interstellar space $B_{\rm ISM}$.
Note that the diffusion coefficient depends on the magnetic-field strength and the degree of turbulence; for simplicity, we do not treat the diffusion coefficient as a function of $B_{\rm ISM}$ here.

Under the assumption that the cloud is a sphere with a radius of $R_{\rm cl}$, the CR spectrum in the cloud region is written as follows:
\begin{equation}
\end{equation}
where $d_{\rm cl}$ is the distance to the center of the cloud from the center of the SNR, and $s$ is species ($\rm e$=electron, $\rm p$=proton).
The energy spectrum of CRs in the SNR shell, $N_{s,{\rm shell}}$, is calculated consistently with the distribution function of escaped particles.
By using the normalization equation (\ref{eq:Nesc_normalize}) same as equation (\ref{eq:Nesc}), the spectra at $E<E_{\rm esc}$ of CR protons and electrons in the SNR shell are written as
\begin{equation}
N_{p,{\rm shell}}(E)=N_{\rm esc}(E_{\rm now})\left(\frac{E}{E_{\rm now}}\right)^{-p_{\rm SNR}},
\end{equation}
and
\begin{equation}
N_{e,{\rm shell}}(E)=K_{\rm ep}N_{\rm esc}(E_{\rm now})\left(\frac{E}{E_{\rm now}}\right)^{-p_{\rm SNR}},
\end{equation}
respectively, where $p_{\rm SNR}$ is the index of CRs that have not yet escaped from the SNR shell and is obtained as $p_{\rm SNR} = 2$ from the radio spectrum \citep{Clark1975, Green2019JApA}.

\section{Secondary electrons and positrons in the Cloud} \label{sec:appendix:secondary_SED}

CR protons incident on the cloud from the SNR shell produce not only neutral pions but also charged pions through p-p collisions.
Furthermore, these charged pions eventually decay into electrons, positrons, and neutrinos.
The secondary electrons and positrons emit photons via synchrotron radiation.
In this appendix, we construct a simple one-zone model to estimate the amount of these secondary electrons and positrons in the cloud and the flux of synchrotron radiation emitted.

Let $N_{e, {\rm 2nd}}(t,E)$ represent the energy distribution of the secondary electrons and positrons in the cloud.
The injection rate $(dN_{e, {\rm 2nd}}/dt)_{\rm inj}$ of secondary electrons and positrons into the cloud is determined by the rate of production via p-p collisions, which is obtained using the \verb|AAfrag| package \citep{Koldobskiy2021PhRvD}.
These injected electrons and positrons take approximately the escape timescale $t_{\rm esc}=R_{\rm cl}^2/ 2D_{\rm ISM}$ to diffuse out of the cloud.
Additionally, the magnetic field in the cloud causes the electrons and positrons to lose energy over the synchrotron cooling timescale $t_{\rm syn}=(Q_{\rm syn} E)^{-1}$, where $Q_{\rm syn}$ is evaluated using $B_{\rm cl}$.
Considering these processes, we model the time evolution of the energy spectrum $N_{e, {\rm 2nd}}(t,E)$ of electrons and positrons in the cloud as follows:
\begin{equation}
\frac{\partial N_{e, {\rm 2nd}}}{\partial t}=\left(\frac{d N_{e, {\rm 2nd}}}{dt}\right)_{\rm inj}-\frac{N_{e, {\rm 2nd}}}{t_{\rm esc}(E)}-\frac{N_{e, {\rm 2nd}}}{t_{\rm syn}(E)}.
\end{equation}
By integrating this equation from $t=t_{\rm sedov}$ to $t_{\rm now}$, we can obtain the spectra of secondary electrons and positrons in the cloud at present.
Based on the obtained spectra, we use \textit{naima} \citep{Zabalza2015ICRC} to calculate the synchrotron radiation flux, following the same procedure used for primary electrons from the SNR shell (see Sect. \ref{section:modeling}).

The synchrotron flux due to secondary electrons obtained with the fiducial parameters summarized in Table~\ref{tab:modelpara} is $1.7 \times 10^{-14}~\rm{erg\,cm^{-2}s^{-1}}$ at the spectral peak of ${\sim}0.3$~keV, which is about an order of magnitude smaller than that of primary electrons, as shown in Fig.~\ref{fig:modeling:best}.

\bibliography{references}{}
\bibliographystyle{aasjournal}

\end{document}